\documentclass[journal,twoside,web]{ieeecolor}
\usepackage{tmi}
\usepackage{cite}
\usepackage{amsmath,amssymb,amsfonts}
\usepackage{graphicx}
\usepackage{textcomp}
\usepackage{multirow}
\usepackage{booktabs}
\usepackage{algorithm}
\usepackage{algpseudocode}
\usepackage{multicol}
\usepackage{balance}
\usepackage{xfrac}

\usepackage[pagebackref=true,breaklinks=true,colorlinks=true,bookmarks=false,citecolor=blue,urlcolor=blue,linkcolor=blue]{hyperref}

\renewcommand{\tilde}{\widetilde}

\DeclareMathOperator{\R}{{\mathbb R}}

\DeclareMathOperator{\D}{{\mathcal D}}

\DeclareMathOperator{\cL}{{\mathcal L}}

\DeclareMathOperator{\cP}{{\mathcal P}}
\DeclareMathOperator{\X}{{\mathcal X}}
\DeclareMathOperator{\Y}{{\mathcal Y}}

\DeclareMathOperator*{\argmin}{arg\,min}

\newcommand{\N}{\mathcal{N}}

\newcommand{\black}[1]{\textcolor{black}{#1}}

\def\BibTeX{{\rm B\kern-.05em{\sc i\kern-.025em b}\kern-.08em
    T\kern-.1667em\lower.7ex\hbox{E}\kern-.125emX}}
\begin{document}
\title{IOP-FL: Inside-Outside Personalization for \\ Federated Medical Image Segmentation}
\author{Meirui Jiang~\IEEEmembership{Student Member,~IEEE}, Hongzheng Yang~\IEEEmembership{Student Member,~IEEE}, \\Chen Cheng~\IEEEmembership{Member,~IEEE}, Qi Dou~\IEEEmembership{Member,~IEEE}
\thanks{Manuscript submitted on 13 April 2022; accepted 20 March 2023. This research was supported by National Natural Science Foundation of China (Project No. 62201485), the Research Grants Council of the Hong Kong Special Administrative Region, China (Project No. T45-401/22-N), and the Hong Kong Innovation and Technology Commission (Project No. ITS/238/21).
\emph{Corresponding author: Qi Dou.}}
\thanks{Meirui Jiang, Hongzheng Yang, Cheng Chen, and Qi Dou are with the Department of Computer Science and Engineering, The Chinese University of Hong Kong, Hong Kong, China.
(Emails: \{mrjiang, hzyang22, cchen, qdou\}@cse.cuhk.edu.hk)}
}

\maketitle

\begin{abstract}
Federated learning (FL) allows multiple medical institutions to collaboratively learn a global model without centralizing client data. It is difficult, if possible at all, for such a global model to commonly achieve optimal performance for each individual client, due to the heterogeneity of medical images from various scanners and patient demographics. This problem becomes even more significant when deploying the global model to unseen clients outside the FL with unseen distributions not presented during federated training. To optimize the prediction accuracy of each individual client for medical imaging tasks, we propose a novel unified framework for both \textit{Inside and Outside model Personalization in FL} (IOP-FL). Our inside personalization uses a lightweight gradient-based approach that exploits the local adapted model for each client, by accumulating both the global gradients for common knowledge and the local gradients for client-specific optimization. Moreover, and importantly, the obtained local personalized models and the global model can form a diverse and informative routing space to personalize an adapted model for outside FL clients. Hence, we design a new test-time routing scheme using the consistency loss with a shape constraint to dynamically incorporate the models, given the distribution information conveyed by the test data. Our extensive experimental results on two medical image segmentation tasks present significant improvements over SOTA methods on both inside and outside personalization, demonstrating the potential of our IOP-FL scheme for clinical practice. Code is available at \href{https://github.com/med-air/IOP-FL}{https://github.com/med-air/IOP-FL}.
\end{abstract}


\section{Introduction}
\label{sec:introduction}
\begin{figure}[t]
\centering
\includegraphics[width=1\columnwidth]{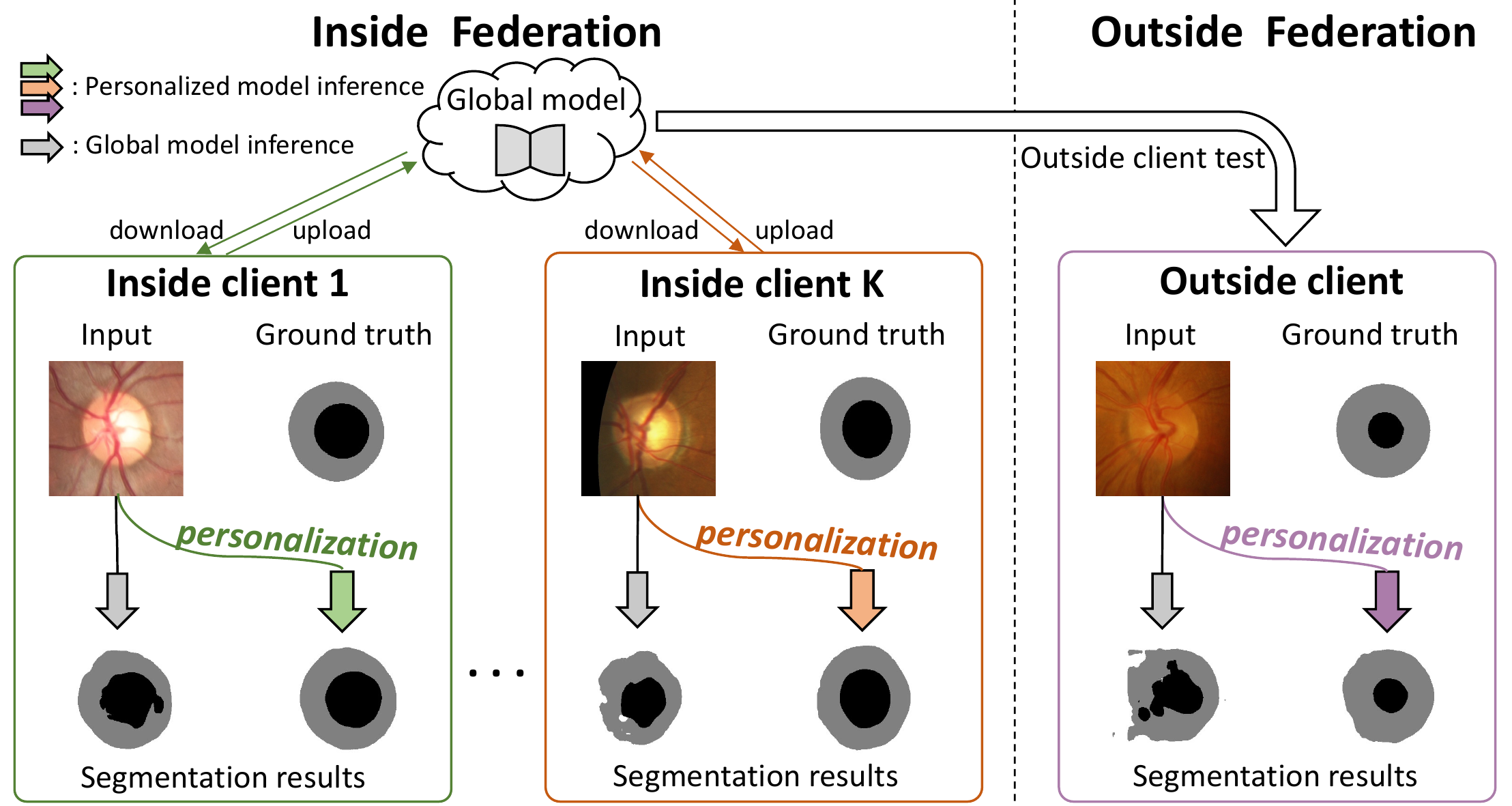} 
\vspace{-5mm}
\caption{Illustration of the motivation for inside-outside personalization in FL. The global model may not be optimal (gray arrows) for all clients due to data heterogeneity. Our method can personalize the global model to yield the best possible performance for each training and testing client.
}
\vspace{-5mm}
\label{fig:coverpage}
\end{figure}

Federated learning (FL) in medical image analysis has been an increasingly important topic, owning to its advantage of jointly learning a global model on many clients in a decentralized way~\cite{rieke2020future,silva2019federated,sheller2020federated,roth2020federated,fedsim,li2020multi,yeganeh2020inverse,dayan2021federated,ju2020federated,dou2021federated,jiang2021harmofl,liu2021federated}. 
However, such a global model is unlikely to be optimal for each individual FL client (see Fig.~\ref{fig:coverpage}), due to various data distributions associated with different scanners, protocols, patient demographics at the clients~\cite{tan2022toward,wu2020personalized,kulkarni2020survey,kairouz2021advances,hsieh2020non,xu2021federated,fedbn}. 
Thus, instead of training a global model that is to be commonly used everywhere, we argue it is crucial to personalize FL models, in order to optimize prediction accuracy at each client for critical medical applications.
Moreover, to make the best use of the federated training effort, it is good if the federated model can also perform well on clients outside federation in the wild.
Unfortunately, due to the data distribution shift, the global model may encounter even more severe performance degradation on unseen testing clients.
In these regards, exploring FL frameworks equipped with comprehensive personalization strategies for both inside and outside clients is highly demanded.

Existing works relevant to this challenging problem have been mainly on model personalization for inside FL clients.
Some approaches incorporate extra learning paradigms to balance local and global information for each local client model, such as multi-task learning~\cite{dinh2021fedu,smith2017federated} and meta learning~\cite{pfedme,fallah2020personalized}.
These methods would complicate the federated training, e.g., multi-task learning has to add auxiliary tasks and solve them simultaneously.
Some recent works aim to specify part of network parameters for personalization~\cite{fedbabu,fedrep,fedbn,fedper}, such as the client-specific classification heads~\cite{fedbabu,fedrep}.
However, it is non-trivial to identify and separate model parameters between the shareable and personalizable subsets. 
Moreover, all these previous personalization approaches have to be performed during the federated training, and are hardly applicable to outside FL clients during the model deployment. 

For outside FL model personalization, the key challenge is unavailability of original data distributions of FL training, i.e., only the learned FL model and test data with new distributions are given during testing.
Furthermore, for the sake of wider applicability, we consider practical situations that no labels are provided by outside clients and the personalization needs to be achieved at test time. 
Recently, some test-time adaptation methods~\cite{ttt,wang2021tent} aim to use the inference sample as a hint of the new data distribution so as to adapt certain model layers for better performance. 
However, directly combining test-time adaptation methods with personalized FL methods in an ad-hoc way can be less effective. One reason for this is that test-time adaptation methods are designed for a single model, making it difficult to determine whether a personalized internal model or the global model is better for adaptation.

In this paper, we propose a novel and unified framework to achieve both \textit{\textbf{I}nside and \textbf{O}utside \textbf{P}ersonalization in \textbf{FL} (\textbf{IOP-FL})}.
For the inside personalization, we design a gradient-based method by calculating a local adapted model as the personalization for each inside client. 
Intuitively, the global gradients aggregated from all clients contain the general information, and the local gradients based on each client's own data represent the local data distribution. 
Our local adapted model dynamically accumulates both types of gradients to make each personalized model optimize towards the client-specific direction without diverging from the agreed common knowledge.
Moreover, for outside personalization in IOP-FL, since the data distribution of an outside client is not presented during federated training, we consider it is important to construct a more diverse and informative parameter space as a good basis to personalize an adapted model at the test time.
Such a space can be well-composed with our obtained various local personalized models representing different data distributions, as well as the common global model for general patterns. 
In this regard, we design a coefficient matrix to dynamically combine the local personalized models and the global model in a layer-wise manner, which is optimized by a new test-time routing scheme to adapt to the unseen distributions of inference data.
The test-time personalization is driven by a consistency regularization with a shape constraint for segmentation tasks in an unsupervised way. We have conducted comprehensive experiments on two medical image segmentation tasks (i.e., prostate MRI segmentation and retinal fundus image segmentation) to validate our approach. 
Our main contributions are highlighted in the following:

\begin{itemize}
    \item We propose a novel FL framework for the first time to simultaneously achieve inside-outside personalization for medical image analysis with heterogeneous data.

    \item We develop a gradient-based personalization method by calculating local adapted models with the global and local gradients, which can effectively and efficiently yield personalized models for inside FL clients. 
    
    \item We design a test-time personalization scheme which dynamically combines the various local personalized models and the global model in an unsupervised way to obtain personalized models for outside FL clients.
    
    \item On two medical image datasets, our IOP-FL framework achieves superior performance on both inside and outside clients over current state-of-the-art (SOTA) methods.
    
\end{itemize}

\section{Related Work}
\subsection{Model Personalization for Inside Federation Clients}
Personalized federated learning aims to learn a local model that is personalized to the distribution of each client.
A variety of approaches have been proposed to achieve model personalization for inside FL clients by using local fine-tuning~\cite{wang2019federated}, multi-task learning~\cite{dinh2021fedu,smith2017federated}, clustering~\cite{ghosh2020efficient,sattler2020clustered}, transfer learning~\cite{yu2020salvaging}, knowledge distillation~\cite{li2019fedmd} and meta-learning~\cite{pfedme,fallah2020personalized}.
The recent state-of-the-art methods try to specify the entire network into the shared global parameters and the unique local parameters~\cite{fedbabu,fedrep,fedbn,fedper}. For example, FedRep~\cite{fedrep} shares the feature extractor and personalizes a classification head for each client. In FedBN~\cite{fedbn}, the batch normalization layers are kept locally, which can be used for the purpose of personalization. Recently, a few early attempts have been made to study personalized FL in medical applications. Roth et al.~\cite{roth2021federated} combine FL with an AutoML technique and propose an adaptation scheme to obtain personalized model architectures for each client. Chen et al.~\cite{chen2021personalized} propose a personalized retrogress-resilient framework to achieve a personalized model for each client with improved performance. 
All the previous personalized FL works focus only on the model personalization for clients inside the federation, but we propose a unified FL framework to personalize the model for both inside and outside clients in which the data distributions are different.  

\subsection{Model Generalization to Outside Federation Clients}
There are currently some works aiming to improve the performance of federated models on outside unseen sites. The work FedDG~\cite{liu2021feddg} shares frequency information across clients to improve the global model's generalizability to unseen domains. Another work FedADG~\cite{fedadg} introduces a distribution generator to measure and align different distributions to learn domain-invariant representations.
However, the global model in these methods is not dedicatedly optimized towards the distributions of outside FL clients, thus still cannot take care of the high performance for unseen clients. 
Besides, extra training cost in these methods (e.g., frequency information sharing in FedDG, domain generator in FedADG) needs to be considered.

\begin{figure*}[t!]
\centering
\includegraphics[width=1\textwidth]{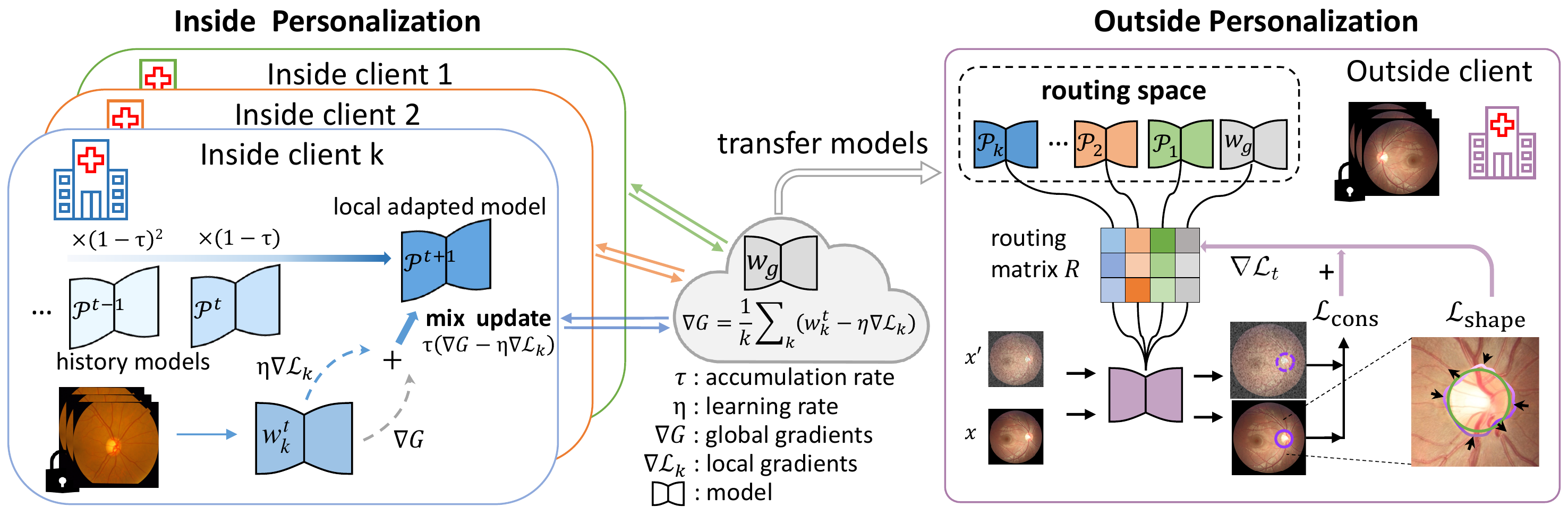}
\vspace{-5mm}
\caption{Overview of our proposed inside-outside personalization FL framework (IOP-FL). For inside clients, each client learns a local adapted model as the inside personalized model, by injecting both global and local gradients. For outside clients, a diverse and informative routing space is constructed with the local personalized models and the global model, from which an outside personalized model is dynamically updated at test time. The test-time routing is driven by the consistency task with a shape constraint loss, using only unlabeled inference data.}
\label{fig:overview}
\end{figure*}

Test-time learning is recently proposed for domain adaptation/generalization to obtain better predictions on new domains by adapting models with the distribution information presented at test time. Test-time-training (TTT)~\cite{ttt} designs a rotation recognition of images as the auxiliary task to adapt the encoder of models. Tent~\cite{wang2021tent} measures and minimizes an entropy loss to conduct the test-time adaptation on the batch normalization layers.
Although these methods are applicable to the model personalization outside federation, their model updating is restricted to a limited parameter space and is not able to fully utilize the diverse personalized FL models of internal clients. 
Instead, our test-time update scheme tries to personalize a new model from a diverse parameter space represented by our learned various local personalized models, which can largely enhance the representation capability at the test-time.
Experimental comparison with these approaches demonstrates the superior performance of our method.

\section{Method}
To achieve model personalization for both inside and outside clients in federated learning, we propose a novel unified FL framework (named IOP-FL) as illustrated in Fig.~\ref{fig:overview}. In this following section, we will first provide the overall formulation of IOP-FL framework. Next, we will describe the local adapted model for the inside personalization. Then, we will introduce the module of test-time routing for the outside personalization.

\subsection{IOP-FL Framework}
\textbf{Preliminaries:}
Let ($\X, \Y$) be the joint image and label space, $\{\D_{1}^s,\cdots,\D_{K}^s\}$ as the set of distributions of $K$ distributed training clients involved in FL, and $\D^o$ be the outside client's distribution\footnote{Here we use one outside client for simplification, this framework can be easily extended to multiple clients.}. We consider the heterogeneous distributions that all samples $X$ are non-iid across clients (e.g., medical images from different hospitals). Specifically, for the $k$-th client, let $\{(x^s_{i,k},y^s_{i,k})\}_{i=1}^{n_k}$ be the training samples drawn independently from a training distribution $\D_k^s$, and $\{x^o_i\}^{n_o}_{i=1}$ be testing samples from $\D^o$ without labels available.

The FL aims to find the best global model parameterized by $w$ by solving the overall empirical risk minimization problem, i.e., 
$\min _{{w}} \frac{1}{K} \sum_{k=1}^{K} 
\frac{1}{n_k}\sum_{i=1}^{n_k} \cL(f_{{w}}(x_{i,k}^{s}), y_{i,k}^{s})$, where $\cL$ is the loss function and $f_{w}$ is the model.
The loss of the client $k$ only depends on its own local data. There is a potential issue of the standard objective. Specifically, the global model may not always achieve the best performance for a given client with a distribution differing from the population significantly. To address this problem, our framework will use personalized objectives to optimize the specific model for each client.

For inside FL clients, we have the objective as follows:
\begin{equation}
\label{eq:obj_inside}
    \min_{\forall k: w_k } \left\{ \frac{1}{K} \sum\nolimits_{k=1}^{K} 
    \mathcal{L}_{k}
    + \frac{\lambda}{K}\sum\nolimits_{k=1}^{K} \text{sim}(w_k, \bar{w})\right\},
\end{equation}
where $\bar{w} = \frac{1}{K}\sum_{k=1}^K w_k$ is the average global model, {$\cL_k$ denotes the local loss} and $\lambda$ indicates the trade-off on the model similarity measured by the function $\text{sim}(\cdot, \cdot)$. 

For outside FL clients, we have the objective as follows:
\begin{equation}
\label{eq:obj_outside}
    \min_{w_o \in \tilde{\mathcal{W}}} \left\{\frac{1}{n_o}\sum\nolimits_{i=1}^{n_o}\mathcal{L}_t(f_{w_o}(x_{i}^{o}))\right\},
\end{equation}
where $\tilde{\mathcal{W}} = R \cdot [w_1,\cdots,w_k, \bar{w}]^{\top}$
is the model weight routing space that incorporates all models obtained after FL training, including the local personalized models and the global model;
the $R\in\R^{L\times(K+1)}$ is the multiplication routing matrix consisting of re-weight coefficient terms $r_k^l$ for the $l$-th layer of the $k$-th model. All terms in the routing matrix are updated by a specifically designed test-time loss $\mathcal{L}_t$ (cf. Eq. (\ref{eq:overall_loss})).

\subsection{Local Adapted Model for Inside Personalization}
A good local personalized model should make full use of the information that all clients contribute to mitigating insufficient local data, and also be optimized w.r.t. each individual client's own distribution. 
With this insight, we propose to calculate the local adapted model, which gathers both the global gradients and local gradients in an accumulated way for inside personalization.  
The global gradients help extract a relatively unbiased description of the general pattern to make each client benefit from the large distributed data and the local gradients help to personalize toward each client's distribution. Benefitting from the congregating of gradients, the local adapted model contains information in the current round, and also keeps tracking of information in previous rounds. This alleviates the influence of client heterogeneity and stabilizes the local personalization.

To step into the specific design of the local adapted model, we start by solving the objective of Eq.~(\ref{eq:obj_inside}), which allows each client to learn a specific model $w_k$ different from each other under a constraint of dissimilarity measurement. This objective has been emerging explored in existing literature~\cite{fedprox,hanzely2020federated,pfedme}, e.g. using the L2-norm to penalize model differences to learn a common global model or designing a new loss term to update the local model. But these methods are sensitive to the choice of $\lambda$ in Eq.~(\ref{eq:obj_inside}). Ideally, the local models are optimized towards different directions regarding the value of $\lambda$. When $\lambda=0$, each client $k$ only calculates local gradients based on samples drawn from $\mathcal{D}_k^s$. 
When considering the limit case $\lambda \rightarrow \infty$, intuitively, this limit should force each local model to be identical, which is equivalent to optimizing solely with global average gradients. In this case, for $\lambda \in (0,\infty)$, the local models should be an interpolation between pure local models and the global model. Such interpolation can gain benefits from both global gradients and local gradients. 

{In FL, the global gradients aggregation typically happens after one local update epoch, i.e., after iterating over all local samples, clients will send local gradients to server for aggregation. From the viewpoint of local client, the local model is optimized using either local gradients during local training, or global aggregated gradients after aggregation. In this regard, the local client training process can be seen as an alternating optimization with either local or global gradients. Let client $k$ has $m_k$ local iteration steps, denote each step gradient as $\nabla\cL_{k,i}$ and global aggregated gradients as $\nabla_G = \sfrac{1}{K}\sum_{k=1}^K \sum_{i=1}^{m_k} \nabla\cL_{k,i}$. Given a model at $i$-th iteration, if it is the last local iteration, i.e., $i=m_k$, then the model at $i+1$ step will be optimized using global gradients, otherwise, the model will be optimized using local gradients.
With the condition of aggregation or not, the process can be formulated as:}

\begin{equation}
\label{eq:localgd}
{
\vspace{-3mm}
    w_{k}^{i+1}=\left\{\begin{array}{ll} w_{k}^{i}-\eta_g \nabla_G & \text{if } aggregation,\\
    w_{k}^{i}-\eta_l \nabla \cL_{k,i} & \text{else},\end{array}\right.
    }
\end{equation}
where $\eta_g$ and $\eta_l$ denote the learning for global aggregation and local update. The process of local model training can be seen as the alternating optimization with either local gradients or overall global gradients. 
{Based on this alternating optimization process, we choose to mix the local gradients and the global gradients as a natural constraint of the dissimilarity measurement $sim$, thereby deriving a local adapted model. By doing this, the local adapted model is updated towards learning a common pattern as well as fitting the local distinct distribution, and we do not need to explicitly specify the $\lambda$. Denote the local adapted model as $\cP_k$. To make the problem well-defined, we specify $\cP_k^{t} = w_k^t$ when the communication round $t=0$. Then the local adapted model of each local client can be calculated as follows: 
\begin{equation}
\label{eq:trajectory}
   {
   \cP_k^{t+1} 
   =  (1-\tau) \cP_k^t - \tau \left(\eta_l \nabla_{L}^k + \eta_g \nabla_G \right),
   }
\end{equation}
where $\tau$ is the accumulation rate to congregate history information and stabilize training. Local client gradients $\nabla_{L}^k = \sum_{i=1}^{m_k} \nabla\cL_{k,i}$ is the sum of $m_k$ local iterations. After communication with server, the local gradients are further calculated and mixed with global gradients to jointly optimize the local adapted model. In our implementation, we experimentally set the same rate of $0.9$ for all clients to help align the history knowledge accumulation status. We further study the effects on different choices of $\tau$ in our experiments.} For the local gradient descent, in practice, we iteratively conduct the stochastic gradient descent on all sampled local training data. The aggregation is naturally performed in the federated learning after certain local training epochs, the calculation of the local adapted model based on global and local gradients does not require extra information.

\subsection{Test-time Routing for Outside Personalization}
With the obtained models during the inside personalization, we further study the personalization on new testing data outside the federation, which is more challenging due to the unseen distribution and unavailable labels. 
Unfortunately, either directly deploying a single model or a straightforward ensemble of models can not deliver good performance due to the unseen distributions of the test data. Although current test-time adaptation methods are applicable, they are only for a single model and are mainly restricted to updating a limited subset of parameters. A smart way to utilize multiple models in the personalized FL needs to be proposed. 
Therefore, based on these internal models, we design a test-time routing scheme to overcome these challenges. This test-time routing aims to use test data to find optimal re-weight coefficients to aggregate all learnable layers from all models, thus incorporating both training knowledge and test data information flexibly.

The objective of the test-time routing is shown in Eq.~(\ref{eq:obj_outside}).
We aim to obtain a new personalized model $w_o$ from all training stage obtained models, we call the collection of these models as well as the corresponding re-weight coefficients as the routing space, which is denoted by $\tilde{\mathcal{W}}$. By minimizing a specifically designed unsupervised loss on the test data from the outside client, we can dynamically update these coefficients to obtain a new outside personalized model from the routing space. The details for routing space construction and test-time routing are described below.

\begin{figure}[t!]
\centering
\includegraphics[width=0.99\columnwidth]{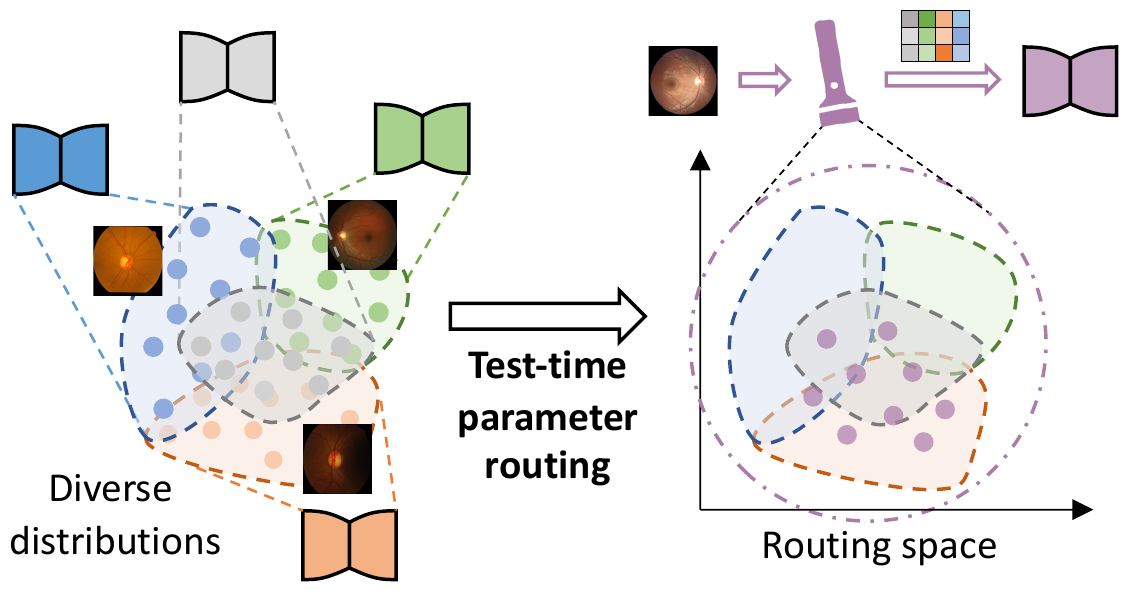}
\vspace{-5mm}
\caption{The illustration of the test-time parameter routing. All trained models present diverse distribution information and the test-time routing aggregates the plentiful parameters to fit the new test data.}
\label{fig:routing_space}
\end{figure}

\textbf{Routing space construction.} 
To unleash the potential of test-time routing, we construct the routing space by considering all personalized models as well as the global model. The global model contains strong common knowledge of the general pattern and each personalized model represents its local distribution. By combining these diverse models in the routing space, the representation capacity is largely improved. 
Furthermore, different from traditional ensemble methods which simply aggregate outputs at the model level, to diversify the feature extraction procedure, we consider a layer-wise manner. {For example, for the $l$-th convolutional layer with parameter $W_k^l$ from the $k$-th model, given sample $x_i^o\sim\D^o$, we aim to find the new weight $W_o^l$ from a space that is linearly combined with {$\{W_k\}_{k=1}^{K+1}$}, where $\{W_k\}_{k=1}^K$ are layers of inside personalized models and $W_{K+1}$ is the layer from the global model:}
\begin{equation}
\label{eq:mix_conv}
{
W_o^l \times x_i^o = \left(r_{1}^l \cdot W_{1}^l+\ldots+r_{K+1}^l \cdot W_{K+1}^l \right) \times x_i^o,}
\end{equation}
{where $r_k^l$ is the re-weight coefficient for the linear combination. We use the test data to dynamically learn the coefficient $r_k^l$ for each learnable layer. The coefficients of $L$-layer models from $K$ clients form the routing matrix $R\in\R^{L\times(K+1)}$, i.e., $R(k,l)$ is the re-weight coefficient $r_k^l$ for the $l$-th layer of the $k$-th model.
We consider all convolutional layers in the neural network. For the batch normalization layer, since it mainly serves for feature normalization by calculating the mean and variance of features, we choose to directly calculate the mean and variance from the test image feature instead of combining all layers.
As shown in Fig.~\ref{fig:routing_space}, given all model weights, we aim to optimize the routing matrix $R$ to calculate the outside personalized model weight $w_o$ from the routing space $\tilde{\mathcal{W}} = R \cdot [\cP_1,\cdots, \cP_K, w_g]^{\top}$.
Note that we only optimize $R$ but keep existing model weights fixed. 
The routing matrix $R$ is the core to present the character of the dynamic.} Intuitively, given an outside client, the data distribution of this client could be similar to one of the inside clients or lies in their mixed distributions. For the first case, each column vector in the matrix would degrade closely to the one-hot format to fit this similar distribution. For another case, the $R$ coordinates all candidates in the model set to generate better parameters that fit test data distribution as closely as possible.
In the following, we will illustrate the design of losses to optimize the coefficients in the routing matrix.

\textbf{Test-time routing process.}  
In this section, we focus on how to combine existing models to obtain a high-performance personalized model. 
The key is to bridge existing models and our desired personalized model via the routing matrix $R$, where this matrix should be optimized regarding the test data distribution. By denoting $\psi$ as the adaptive average pooling, we aim to learn each coefficient $r_k^l$ in $R$ as follows:
\begin{equation}
\label{eq:routing}
r_k^l =1 / (1 + exp({-\psi(h^l) \cdot \theta_k^l})),
\end{equation}
where $h^l$ is the input for the $l$-th layer (for the first layer is the input image and for later layers are feature maps), $\theta_k^l$ is a multi-layer linear network that maps the pooled inputs to calculate the re-weight coefficient. By doing this, each coefficient in the matrix $R$ is determined by the input and the corresponding learnable network $\theta_k^l$. We can dynamically fuse the models to fit the given data from the outside client by optimizing the routing matrix via our designed losses. In specific, we design a consistency regularization to capture the input image feature distribution. Given a test image $x_i^o$, we generate a random Gaussian-like noise $\epsilon\sim \N(0, (\frac{1}{2})^2)$ as a perturbation and add it to image. Denote the output prediction map of $x_i^o$ and $x_i^o+\epsilon$ as $z\in \R^{L\times H \times C}$ and $z^\prime \in \R^{L\times H \times C}$ respectively. We define the consistency loss $\mathcal{L}_{cons}$ as below:
\begin{equation}
\label{eq:consistency}
\mathcal{L}_{cons}=\frac{1}{LH}\sum\nolimits_{i=1}^{LH} \left\|{z_i- z_{i}^\prime}\right\|_2^2.
\end{equation}
Note that this consistency regularization does not need any label information of test data. 
Furthermore, it is observed that for segmentation tasks, the main source of performance drop at unseen clients is the incomplete shape. Concerning this challenge, we further propose a shape-relevant constraint to preserve the regular shape of segmentation masks, by making the model predictions within certain limits to be more consistent.
Denote $V_{(p,q)}^c$ as the probability vector of the segmentation prediction of class $c$ at the position $(p,q)$ and $Q_{(p,q)}$ as the set of prediction probabilities centered at $(p,q)$. We define the distance $d_{(p,q)}$ in the following:
\begin{equation}
\label{eq:distance}
d_{(p,q)} \!=\! \sum\nolimits_{c}(\max_{(p^{\prime},q^{\prime})\in Q_{(p,q)}} V^c_{(p^{\prime},q^{\prime})} - \min_{(p^{\prime},q^{\prime})\in Q_{(p,q)}} V^c_{(p^{\prime},q^{\prime})}),
\end{equation}
where $Q_{(p,q)}= \{(p\pm l, q\pm l^\prime)  |  l,l^\prime\in[d]\}$ and $d$ is a neighbor range radius. {We then design an unsupervised shape constraint by minimizing the distance over all prediction probabilities the entropy (uncertainty) of predictions as below:}

\floatname{algorithm}{Algorithm}
\begin{algorithm}[tb]
\caption{IOP-FL algorithm}
\label{alg:algorithm}
\textbf{Input}: communication rounds $T$, number of clients $K$,  local iteration steps $\{m_k\}_{k=1}^K$, local learning rate $\eta_l$, global learning rate $\eta_g$, accumulation rate $\tau$. \\
\textbf{Output}: {global model $w_{g}$, inside local personalized models $\{\mathcal{P}_{k}\}_{k=1}^{K}$, outside personalized model $w_o$.}\\
\textbf{Inside personalization:}
\begin{algorithmic}[1] 
\State Initialize server model $w_g^{(0)}$, local personalized models $\{\cP_k^{(0)}\}_{k=1}^K = w_g^{(0)}$.
\For{$t=0,1,\ldots, T-1$}
\For{\textit{Client} $k=1,\ldots, K$ \textbf{in parallel}} 
\State receive global gradients $\nabla_G$
\State $\nabla_L^k = \sum_{i=1}^{m_k} \nabla\cL_{k,i}$
\State $\cP_{k}^{t+1} \leftarrow  (1-\tau) \cP_k^t - \tau \left(\eta_l\nabla_L^k + \eta_g\nabla_G \right)$ \Comment{Eq.(\ref{eq:trajectory})}
\State \textbf{return} $\nabla_L^k$
\EndFor
\State $\nabla_G = \frac{1}{K}\sum_{k=1}^K \nabla_L^k$ \Comment{server aggregate}
\State $w_g^{t+1} \leftarrow w_g^t - \nabla_G$ \Comment{global model update}
\EndFor 
\State \textbf{return} $w_g$, $\{\cP_k\}_{k=1}^K$ 
\end{algorithmic}
\textbf{Outside personalization:}
\begin{algorithmic}[1] 
\State Initialize $R(k,l)=\sfrac{1}{(K+1)}$ for all $k=[K+1], l=[L]$, build routing space $\tilde{\mathcal{W}} = R\cdot [\cP_1, \ldots, \cP_K, w_g]^{\top}$
\For{samples $i=1,\ldots,n_o$}
\State $\bar{y}_i^o = f_{w_o} (x_i^o)$ \Comment{model inference}
\State $R = \argmin_{R} \mathcal{L}_t$ \Comment{test-time update, Eq.(\ref{eq:overall_loss})}
\State $w_o = R\cdot [\cP_1, \ldots, \cP_K, w_g]^{\top}$
\EndFor
\State \textbf{return} $w_o$
\end{algorithmic}
\end{algorithm}

\begin{table*}[!t]
    \renewcommand\arraystretch{1.2}
    \centering
        \caption{\small{Comparison of \textbf{inside FL} results on Dice scores with image segmentation datasets of the \\ prostate MRI images and retinal fundus images.}}
        \centering
        \vspace{-2mm}
        {
        \scalebox{1.}{
        \setlength\tabcolsep{3pt}
        \begin{tabular}{c|ccccccc||ccccc|ccccc}
            \hline
            \hline
             \multirow{2}{*}{Task}&\multicolumn{7}{c||}{\multirow{2}{*}{Prostate MRI Segmentation}}&\multicolumn{10}{c}{Retinal Fundus Image Segmentation} \\\cline{9-18}
             &&&&&&&&\multicolumn{5}{c|}{Optic Disc} &\multicolumn{5}{c}{Optic Cup}
              \\\cline{0-17}
            Inside Client &A & B & C & D & E & F & Avg.
            & A & B & C &D & Avg. &A & B & C &D & Avg. \\
            \hline
            \hline
           \multirow{1.5}{*}{pFedMe (NeruIPS 2020)~\cite{pfedme}} 
           &92.70 &86.91 &92.52 &84.27 &90.25 &93.77 &90.07 
           &95.62 &85.17 &94.46 &93.36 &92.15 
           &83.40 &70.68 &85.16 &85.51 &81.19
           \\ [-1.3ex]
        &\scriptsize{(0.38)} &\scriptsize{(0.70)}&\scriptsize{(0.16)}
        &\scriptsize{(0.42)}&\scriptsize{(0.25)} &\scriptsize{(0.15)}&\scriptsize{(0.40)} &\scriptsize{(0.22)}&\scriptsize{(0.40)} &\scriptsize{(0.37)}
        &\scriptsize{(0.24)}&\scriptsize{(0.21)} &\scriptsize{(0.32)} &\scriptsize{(0.37)}&\scriptsize{(0.36)} &\scriptsize{(0.20)}&\scriptsize{(0.07)}
           \\ [-0.5ex]
           \multirow{1.5}{*}{FedRep (ICML 2021)~\cite{fedrep}}
           &92.97 &87.11 &92.69 &85.71 &90.12 &\textbf{93.84} &90.41
           &95.41 &86.83 &94.59 &93.32 &92.54 
           &83.49&70.96 & 86.02&85.81 &81.57  \\  [-1.3ex]
        &\scriptsize{(0.66)} &\scriptsize{(0.77)}&\scriptsize{(0.59)}
        &\scriptsize{(0.36)}&\scriptsize{(0.42)} &\scriptsize{(0.74)}&\scriptsize{(0.65)} &\scriptsize{(0.39)}&\scriptsize{(0.25)} &\scriptsize{(0.26)}
        &\scriptsize{(0.30)}&\scriptsize{(0.21)} &\scriptsize{(1.00)} &\scriptsize{(0.68)}&\scriptsize{(0.75)} &\scriptsize{(0.33)}&\scriptsize{(0.54)}\\[-0.5ex]
        \multirow{1.5}{*}{FedBN (ICLR 2021)~\cite{fedbn}} 
        &92.79	&87.61	&93.40	&84.22	&89.96	&93.28	&90.21
        &95.18	&84.82	&95.03	&94.80 &92.46 &82.55 &70.83	&85.51	&86.88 &81.44
 \\  [-1.3ex]
        &\scriptsize{(0.22)} &\scriptsize{(0.10)}&\scriptsize{(0.50)}
        &\scriptsize{(0.93)}&\scriptsize{(0.43)} &\scriptsize{(0.35)}&\scriptsize{(0.34)} &\scriptsize{(0.55)}&\scriptsize{(0.81)} &\scriptsize{(0.79)}
        &\scriptsize{(0.36)}&\scriptsize{(0.46)} &\scriptsize{(0.45)} &\scriptsize{(0.34)}&\scriptsize{(0.25)} &\scriptsize{(0.83)}&\scriptsize{(0.42)}\\
        \hline
        \multirow{1.5}{*}{Single-site}
        &88.04 & 91.80 & 90.38 & 84.89 & 91.43 & 80.49 & 87.84
        &93.34 & 79.52 & 92.21 & 92.44 & 89.38
        &79.39 & 68.87 & 83.06 & 81.51 & 78.21
         \\[-1.3ex]
        &\scriptsize{(0.85)} &\scriptsize{(0.52)}&\scriptsize{(0.17)}
        &\scriptsize{(0.32)}&\scriptsize{(0.41)} &\scriptsize{(0.10)}&\scriptsize{(0.82)} &\scriptsize{(0.45)}&\scriptsize{(0.10)} &\scriptsize{(0.12)}
        &\scriptsize{(0.20)}&\scriptsize{(0.12)} &\scriptsize{(0.37)} &\scriptsize{(0.54)}&\scriptsize{(0.33)} &\scriptsize{(0.66)}&\scriptsize{(0.03)}\\
        [-0.5ex]
        \multirow{1.5}{*}{FedAvg\cite{mcmahan2017communication}} 
        &92.29 &85.92 &92.24 &82.03  &90.41 &92.59 &89.25 
        
        &95.32 &82.64 &94.56 &93.37 &91.48 &82.36 &70.92 &85.27 &84.86 &80.85
         \\[-1.3ex]
        &\scriptsize{(0.45)} &\scriptsize{(0.65)}&\scriptsize{(0.22)}
        &\scriptsize{(0.44)}&\scriptsize{(0.37)} &\scriptsize{(0.47)}&\scriptsize{(0.28)} &\scriptsize{(0.37)}&\scriptsize{(0.17)} &\scriptsize{(0.08)}
        &\scriptsize{(0.49)}&\scriptsize{(0.23)} &\scriptsize{(0.19)} &\scriptsize{(0.64)}&\scriptsize{(0.21)} &\scriptsize{(0.65)}&\scriptsize{(0.25)}\\
        [-0.5ex]
        \multirow{1.5}{*}{\textbf{IOP-FL(Ours)}}
        &\textbf{93.36} & \textbf{92.18} &\textbf{94.71}    &\textbf{86.37} 
        &\textbf{91.44} &{93.24} &\textbf{91.88} 
    
        & \textbf{96.27} &\textbf{89.75} &\textbf{95.71}
      &\textbf{95.36} &\textbf{94.27} &\textbf{85.32}
    &\textbf{72.53} &\textbf{86.41} &\textbf{87.48} 
     &\textbf{82.94}\\[-1.3ex]
        &\scriptsize{(0.68)} &\scriptsize{(0.87)}&\scriptsize{(0.49)}
        &\scriptsize{(0.82)}&\scriptsize{(0.94)} &\scriptsize{(0.60)}&\scriptsize{(0.55)} &\scriptsize{(0.52)}&\scriptsize{(0.16)} &\scriptsize{(0.62)}
        &\scriptsize{(0.08)}&\scriptsize{(0.09)} &\scriptsize{(0.24)} &\scriptsize{(0.28)}&\scriptsize{(0.31)} &\scriptsize{(0.21)}&\scriptsize{(0.04)}\\
           \hline
           \hline
        \end{tabular}
    }}

    \label{table:inside_all}
\end{table*}

\begin{table*}[!t]
    \renewcommand\arraystretch{1.2}
    \centering
        \caption{\small{Comparison of \textbf{outside FL} results on Dice scores with image segmentation datasets of the \\ prostate MRI images and retinal fundus images.}}
        \centering
        \vspace{-2mm}
        {
        \scalebox{1.}{
        \setlength\tabcolsep{2.5pt}
         \begin{tabular}{c|ccccccc||ccccc|ccccc}
          \hline
            \hline
             \multirow{2}{*}{Task}&\multicolumn{7}{c||}{\multirow{2}{*}{Prostate MRI Segmentation}}&\multicolumn{10}{c}{Retinal Fundus Image Segmentation} \\\cline{9-18}
             &&&&&&&&\multicolumn{5}{c|}{Optic Disc} &\multicolumn{5}{c}{Optic Cup}\\\cline{0-17}
            Outside Client &A & B & C & D & E & F & Avg.
            & A & B & C &D & Avg. &A & B & C &D & Avg. \\
            \hline
            \hline
          \multirow{1.5}{*}{TTT (ICML 2020)~\cite{ttt}} 
           &89.50 &86.59 &84.80 &87.35 &81.78 &90.32 &86.72
           &94.49 & 86.70 &93.29 & 94.79& 92.32 & 83.19&70.43&83.35 &84.36& 80.33\\
           [-1.3ex]
        &\scriptsize{(0.25)} &\scriptsize{(0.23)}&\scriptsize{(0.26)}
        &\scriptsize{(0.31)}&\scriptsize{(0.66)} &\scriptsize{(0.29)}&\scriptsize{(0.55)} &\scriptsize{(0.48)}&\scriptsize{(0.17)} &\scriptsize{(0.25)}
        &\scriptsize{(0.92)}&\scriptsize{(0.39)} &\scriptsize{(0.21)} &\scriptsize{(0.27)}&\scriptsize{(0.34)} &\scriptsize{(0.25)}&\scriptsize{(0.18)}\\
        [-0.5ex]
           \multirow{1.5}{*}{Tent (ICLR 2021)~\cite{wang2021tent}}
           &90.35 &85.93 &86.06 &88.22 &81.59 &91.76 &87.32 
           &95.26 & 87.46 &93.88 & 94.45& 92.76 & 84.56&71.45 & {84.57} &85.40 &81.50\\
         [-1.3ex]
        &\scriptsize{(0.12)} &\scriptsize{(0.68)}&\scriptsize{(0.13)}
        &\scriptsize{(0.88)}&\scriptsize{(0.19)} &\scriptsize{(0.08)}&\scriptsize{(0.43)} &\scriptsize{(0.57)}&\scriptsize{(0.45)} &\scriptsize{(0.22)}
        &\scriptsize{(0.28)}&\scriptsize{(0.28)} &\scriptsize{(0.27)} &\scriptsize{(0.25)}&\scriptsize{(0.30)} &\scriptsize{(0.20)}&\scriptsize{(0.05)}\\
        [-0.5ex]
           \multirow{1.5}{*}{FedDG (CVPR 2021)~\cite{liu2021feddg}}
           &90.50 &86.71 &85.72 &87.72 &83.70 &90.91 &87.54
           &95.11 & 87.31 &93.59 & 94.39& 92.60 & 84.64&71.35 &83.80 &85.39 &81.29\\[-1.3ex]
        &\scriptsize{(0.28)} &\scriptsize{(0.43)}&\scriptsize{(0.40)}
        &\scriptsize{(0.47)}&\scriptsize{(0.74)} &\scriptsize{(0.38)}&\scriptsize{(0.50)} &\scriptsize{(0.31)}&\scriptsize{(0.33)} &\scriptsize{(0.20)}
        &\scriptsize{(0.10)}&\scriptsize{(0.14)} &\scriptsize{(0.45)} &\scriptsize{(0.67)}&\scriptsize{(0.12)} &\scriptsize{(0.20)}&\scriptsize{(0.16)}\\
           \hline
        \multirow{1.5}{*}{FedAvg\cite{mcmahan2017communication}}
        &88.78 &84.42 &84.33 &86.75  &80.10 &90.45 &85.81
        &92.46 & 86.11 &93.53 & 93.39& 91.37 & 78.57&69.18 &82.07 &83.22 &78.26\\
        [-1.3ex]
        &\scriptsize{(0.47)} &\scriptsize{(0.39)}&\scriptsize{(0.63)}
        &\scriptsize{(0.16)}&\scriptsize{(0.11)} &\scriptsize{(0.52)}&\scriptsize{(0.62)} &\scriptsize{(0.34)}&\scriptsize{(0.09)} &\scriptsize{(0.50)}
        &\scriptsize{(0.15)}&\scriptsize{(0.07)} &\scriptsize{(0.42)} &\scriptsize{(0.57)}&\scriptsize{(0.77)} &\scriptsize{(0.14)}&\scriptsize{(0.41)}\\
        [-0.5ex]
        \multirow{1.5}{*}{Average}
        &87.27 &82.43 &74.41 &84.59 &78.30 &87.34 &82.39
        &93.30 & 84.77 &92.31 &92.63 & 90.75
        &77.67 &61.55 &82.73 &81.21 &75.79 \\
        [-1.3ex]
        &\scriptsize{(0.22)} &\scriptsize{(0.35)}&\scriptsize{(0.48)}
        &\scriptsize{(0.27)}&\scriptsize{(0.35)} &\scriptsize{(0.43)}&\scriptsize{(0.50)} &\scriptsize{(0.31)}&\scriptsize{(0.45)} &\scriptsize{(0.22)}
        &\scriptsize{(0.54)}&\scriptsize{(0.31)} &\scriptsize{(0.41)} &\scriptsize{(0.19)}&\scriptsize{(0.10)} &\scriptsize{(0.21)}&\scriptsize{(0.18)}\\
        [-0.5ex]
        \multirow{1.5}{*}{Ensemble}
        &88.56 &84.39 &76.86 &85.42 &79.65 &89.26 &84.02&94.34 &85.68 &92.76 &92.32 &91.28
        &78.49 &68.52 &82.45 &83.19 &78.16 \\
        [-1.3ex]
        &\scriptsize{(0.28)} &\scriptsize{(0.33)}&\scriptsize{(0.08)}
        &\scriptsize{(0.24)}&\scriptsize{(0.13)} &\scriptsize{(0.14)}&\scriptsize{(0.60)} &\scriptsize{(0.43)}&\scriptsize{(0.51)} &\scriptsize{(0.58)}
        &\scriptsize{(0.40)}&\scriptsize{(0.10)} &\scriptsize{(0.11)} &\scriptsize{(0.32)}&\scriptsize{(0.07)} &\scriptsize{(0.11)}&\scriptsize{(0.10)}\\
        [-0.5ex]
        
        \multirow{1.5}{*}{\textbf{IOP-FL(Ours)}}
        &\textbf{90.52} &\textbf{90.52} &\textbf{88.32} &\textbf{89.39} &\textbf{84.33} &\textbf{92.61} &\textbf{89.28}
        
        &\textbf{96.44} & \textbf{90.20} &\textbf{95.59} & \textbf{96.12} & \textbf{94.59} &\textbf{84.70} & \textbf{72.62} & \textbf{84.63}& \textbf{86.43} &\textbf{82.09} \\[-1.3ex]
        &\scriptsize{(0.33)} &\scriptsize{(0.64)}&\scriptsize{(0.41)}
        &\scriptsize{(0.36)}&\scriptsize{(0.21)} &\scriptsize{(0.23)}&\scriptsize{(0.44)} &\scriptsize{(0.12)}&\scriptsize{(0.26)} &\scriptsize{(0.29)}
        &\scriptsize{(0.08)}&\scriptsize{(0.12)} &\scriptsize{(0.44)} &\scriptsize{(0.24)}&\scriptsize{(0.31)} &\scriptsize{(0.40)}&\scriptsize{(0.21)}\\
           \hline
           \hline
        \end{tabular}
    }}
    
    \label{table:outside_all}
    
\end{table*}

\begin{equation}
\label{eq:shape}
{
\mathcal{L}_{s}  = \sum_{(p,q)}d_{(p,q)}, \quad \cL_{e} = -\sum_{c} V_{(p,q)}^{c} \log V_{(p,q)}^{c}.
}
\end{equation}
{The shape loss improves the segmentation map with a smoother boundary and the entropy loss reduces prediction uncertainty. Finally, we have the overall test-time personalization loss $\mathcal{L}_t$ as follows}:
\begin{equation}
\label{eq:overall_loss}
{
\mathcal{L}_t =\mathcal{L}_{cons}(x_i^o, x_i^o\!+\!\epsilon; R) \!+\! \beta\left(\mathcal{L}_{s}(x_i^o;R) \!+\! \cL_e (x_i^o;R)\right),}
\end{equation}
where $\beta$ is a balancing hyper-parameter, $R$ is the routing matrix consisting of learnable weights to be optimized and $w_o$ is the personalized weights for the outside client. With the complete test-time personalization loss, the routing matrix will be optimized as $\argmin_{R} \cL_t$.
The entire test-time personalization procedure is performed in an unsupervised way and does not affect original training knowledge.

\subsection{Implementation of IOP-FL Framework}
{We put the full algorithm in Algorithm~\ref{alg:algorithm}. It contains inside and outside personalization two stages. In our proposed IOP-FL framework, we use the U-Net~\cite{unet} and the Adam optimizer, the model is trained with a learning rate of $1e^{-3}$ and the batch size is 16.
For inside clients training, we empirically set the accumulation rate $\tau$ as 0.9. }We perform local training for 1 epoch in each communication round and fully train the model for 100 rounds. For outside personalization, we update the routing matrix $R$ for 10 epochs using all test samples. The loss weight $\beta$ is empirically set to 0.01 to balance different loss terms. We use the predictions of the outside personalized model with the lowest test-time unsupervised loss. The framework is implemented with Pytorch 1.7.0, and trained on a server with four NVIDIA TitanXp GPUs.

\section{Experiments}

\subsection{Datasets and Preprocessing}
{We evaluate our approach on the prostate segmentation task with T2-weighted MRI images from 6 different sources (the number of 3D volumes are 12, 12, 30, 30, 13, and 19, respectively)~\cite{liu2020ms,prostate2014evaluation,prostate2015computer,isbi}, and the optic
disc and cup segmentation task with retinal fundus images from 4 different institutions (the number of samples are: 101, 159, 400, and 400, respectively)~\cite{fumero2011fundus,sivaswamy2015fundus,orlando2020fundus}.
We take data from each different source as a single client, constructing the client A to client F for prostate MRI and client A to client D for retinal fundus images.}
For each client, we first randomly split 20\% data as a test set, then the remaining data are further split into 80\% for training and 20\% for validation.
{For data pre-processing, we first resized images to 384 $\times$ 384 for the axial plane of prostate MRI images and retinal fundus images, then we normalized each data individually to zero mean and unit variance.} During training, we adopted data augmentation of random rotation and flipping for all images. The common metric Dice score is employed to evaluate model performance on both tasks.

\begin{figure*}[!h]
\centering
\includegraphics[width=0.99\textwidth]{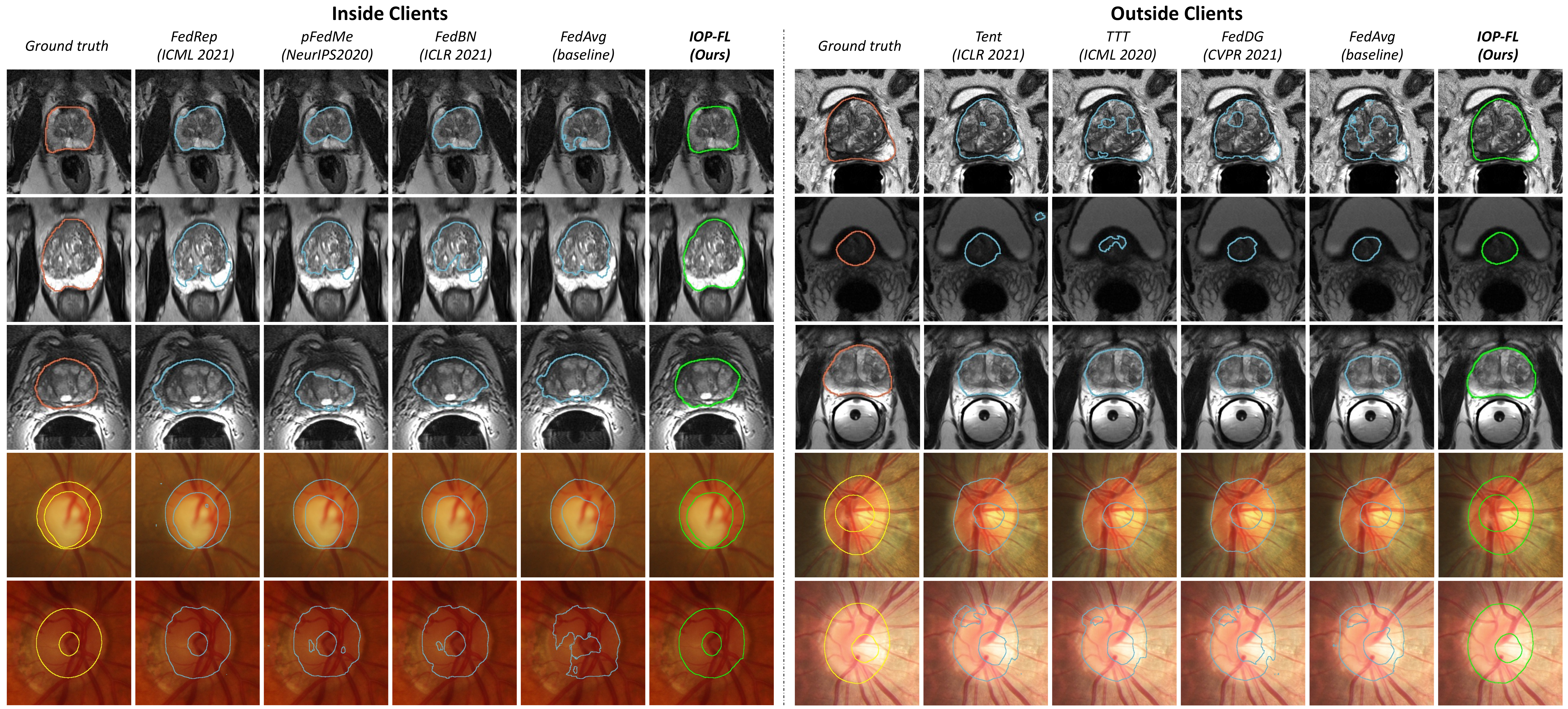}
\vspace{-3mm}
\caption{Qualitative visualization of segmentation results with our inside-outside personalization method and other state-of-the-art methods. The top three rows are for the prostate MRI segmentation and the bottom two rows for retinal fundus image segmentation.}
\label{fig:res_vis}
\vspace{-2mm}
\end{figure*}

\subsection{Comparison with State-of-the-art Personalized FL Methods on Inside Clients}
\textbf{Experimental setting.}
For inside clients comparison, we involve all clients for training and report the performance on each client's test set using the personalized model with the best validation performance.
The performance of our local adapted model is compared with single-site training (i.e., each client trains the model individually), the baseline setting \textbf{FedAvg}~\cite{mcmahan2017communication} (i.e., learn a common global model), and recent SOTA personalized FL methods.
\textbf{FedRep}~\cite{fedrep} (ICML 2021) learns a common representation and a personalized head for each client. \textbf{FedBN}~\cite{fedbn} (ICLR 2021) keeps client-specific batch normalization layers locally.
\textbf{pFedMe}~\cite{pfedme} (NeruIPS 2020) uses the Moreau envelopes as a regularized loss function to pursue clients' own models with different directions.

\textbf{Comparison results.}
Table~\ref{table:inside_all} reports the results on every single client and the average performance for both datasets. All results are in form of average and standard deviation over three independent runs. It can be observed that the FedAvg~\cite{mcmahan2017communication} outperforms single-site training on average, showing the importance for utilizing the multi-source data. On top of this, all personalized FL methods can make further improvement over FedAvg, demonstrating the benefits of model personalization. Compared with these methods, our framework IOP-FL achieves a higher overall performance as well as higher single client performance on 9 out of 10 clients. The improvements attribute to our gradient-based design. Specifically, compared with those partially personalized models, our gradient-based personalization helps the whole model parameters benefit from global and local information, providing a more complete personalization. In addition, the accumulation keeps tracking gradient information, mitigating the instability caused by client heterogeneity during the personalization. 
As a result, our approach achieves consistent improvements over FedAvg on all inside FL clients, with the 2.63\% increase on average Dice for prostate segmentation, as well as 2.79\% and 2.09\% increase on average Dice for optic disc and cup segmentation respectively. 
Besides performance improvements, the left-hand side of Fig.~\ref{fig:res_vis} shows the qualitative visualization of segmentation results in comparison with the baseline FedAvg and other state-of-the-art personalization methods on two tasks. Each row shows an image from a different client and these cases present the heterogeneity. Compared with the ground truth, our method accurately identifies the shape, while other methods sometimes fail to obtain the complete structure.

\begin{figure*}[t]
\centering
\includegraphics[width=1\textwidth]{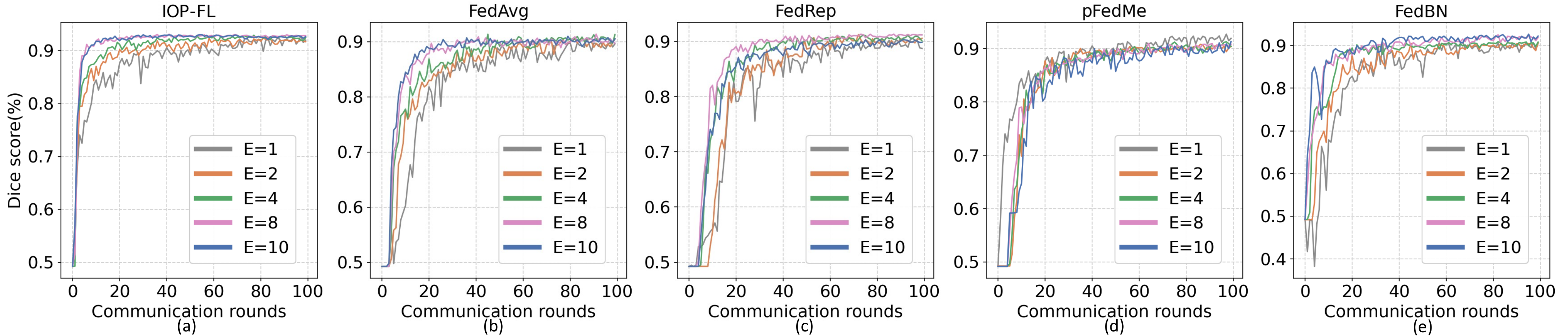}
\vspace{-7mm}
\caption{Training curve of Dice score on inside clients regarding different local update epochs. Each figure represents a distinct method.}
\label{fig:comm_rounds}
\end{figure*}
\subsection{Comparison with State-of-the-art Methods for Personalized FL on Outside Clients}
\textbf{Experimental setting.}
For outside client comparison, we conduct the leave-one-client-out experiment, i.e., each time we leave a client as the outside client and perform the training on routing matrix with unlabeled test data. {Our test-time training will incorporate all training-obtained  models, i.e., the inside personalized models and the global model for the outside personalization.} The results are reported using all data of the outside client. 
Currently, there are very few works on model personalization in FL for outside clients. 
We here compare our method with: 1) a federated domain generalization method \textbf{FedDG}~\cite{liu2021feddg} (CVPR 2021), which shares extra frequency information during federated training to generalize a common global model to outside clients;
2) two test-time adaptation methods to adapt the global model trained by \textbf{FedAvg}~\cite{mcmahan2017communication}, with \textbf{TTT}~\cite{ttt} (ICML 2020) which adapts the encoder by solving image rotation prediction, and \textbf{Tent}~\cite{wang2021tent} (ICLR 2021) which adapts the batch normalization layer by minimizing prediction entropy;
3) the straightforward output ensemble of the obtained local personalized models with \textbf{Average} which averages the model performance of applying each model to the outside client, and \textbf{Ensemble} which obtains the average logits of each model on the outside client for prediction.

\textbf{Comparison results.}
Table~\ref{table:outside_all} shows that FedDG~\cite{liu2021feddg} and test-time learning approaches TTT~\cite{ttt} and Tent~\cite{wang2021tent} can generally improve the model performance over FedAvg on outside clients, but not as effective as our proposed method. This demonstrates the benefits of our test-time personalization from a diverse parameter space that is represented by learned training models. 
We observe that Average and Ensemble approaches obtain inferior performance than FedAvg~\cite{mcmahan2017communication}, showing that simply averaging or ensembling the models can hardly achieve good performance on outside clients which have different data distributions than the inside clients.  
Our proposed test-time routing, instead, can effectively personalize a new model given the distribution information provided by the test data, thus achieving superior performance. 
In specific, we achieve the highest average results on prostate segmentation by increasing $5.26\%$ in Dice over Ensemble. and $3.47\%$ in Dice over FedAvg. 
Superior performance is also obtained with our IOP-FL on retinal fundus images, with the best overall Dice for optic disc/cup segmentation. The segmentation results are shown on the right-hand side of Fig.~\ref{fig:res_vis}, the compared methods may fail to obtain an accurate boundary due to the different unseen data distributions. But with our test-time personalization, our approach shows more accurate segmentation boundaries.

\begin{figure}[t]
\centering
\includegraphics[width=0.99\columnwidth]{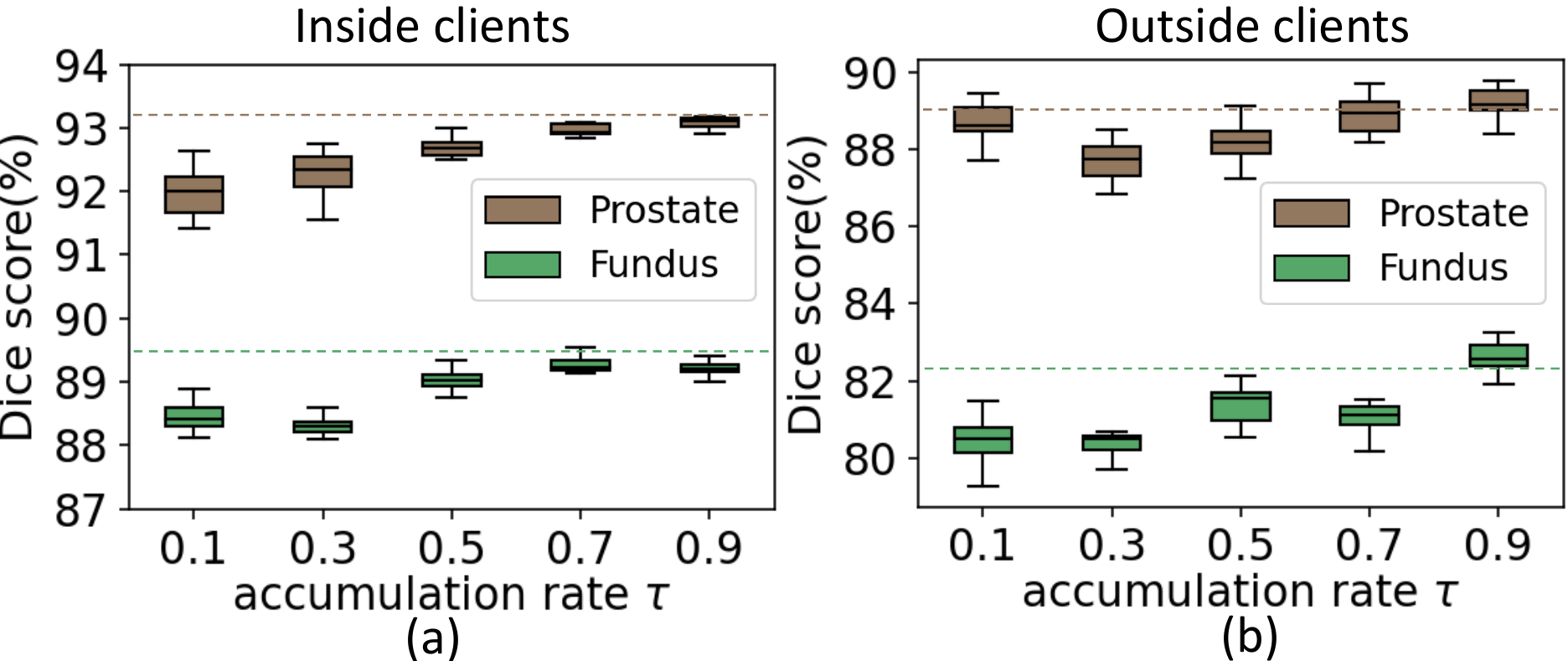} 
\vspace{-3mm}
\caption{Performance of IOP-FL given different accumulation rate $\tau$ on both inside and outside clients.}
\label{fig:vary_tau}
\end{figure}

\begin{figure}[t]
\centering
\includegraphics[width=0.99\columnwidth]{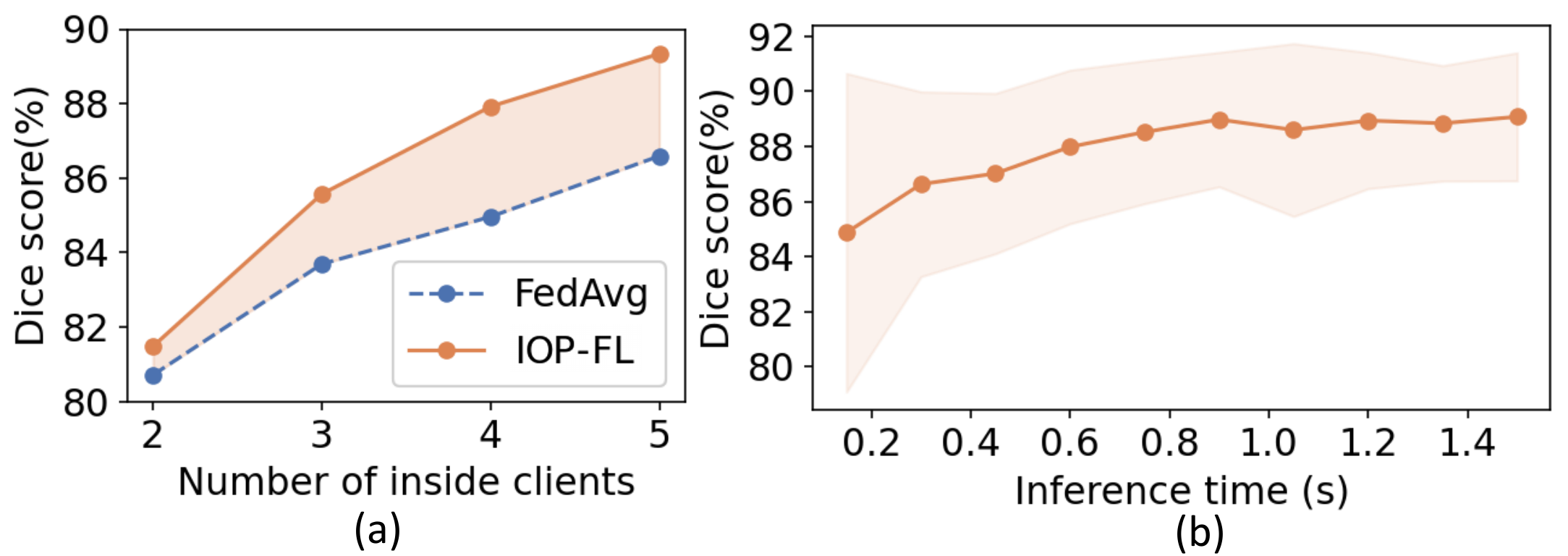} 
\vspace{-3mm}
\caption{The change of Dice score on one outside client for prostate segmentation to investigate (a) the effect of inside participating client number; (b) the relationship between performance and inference time.}
\label{fig:inf_acc_scalable}
\end{figure}

\begin{figure}[t]
\centering
\includegraphics[width=0.98\columnwidth]{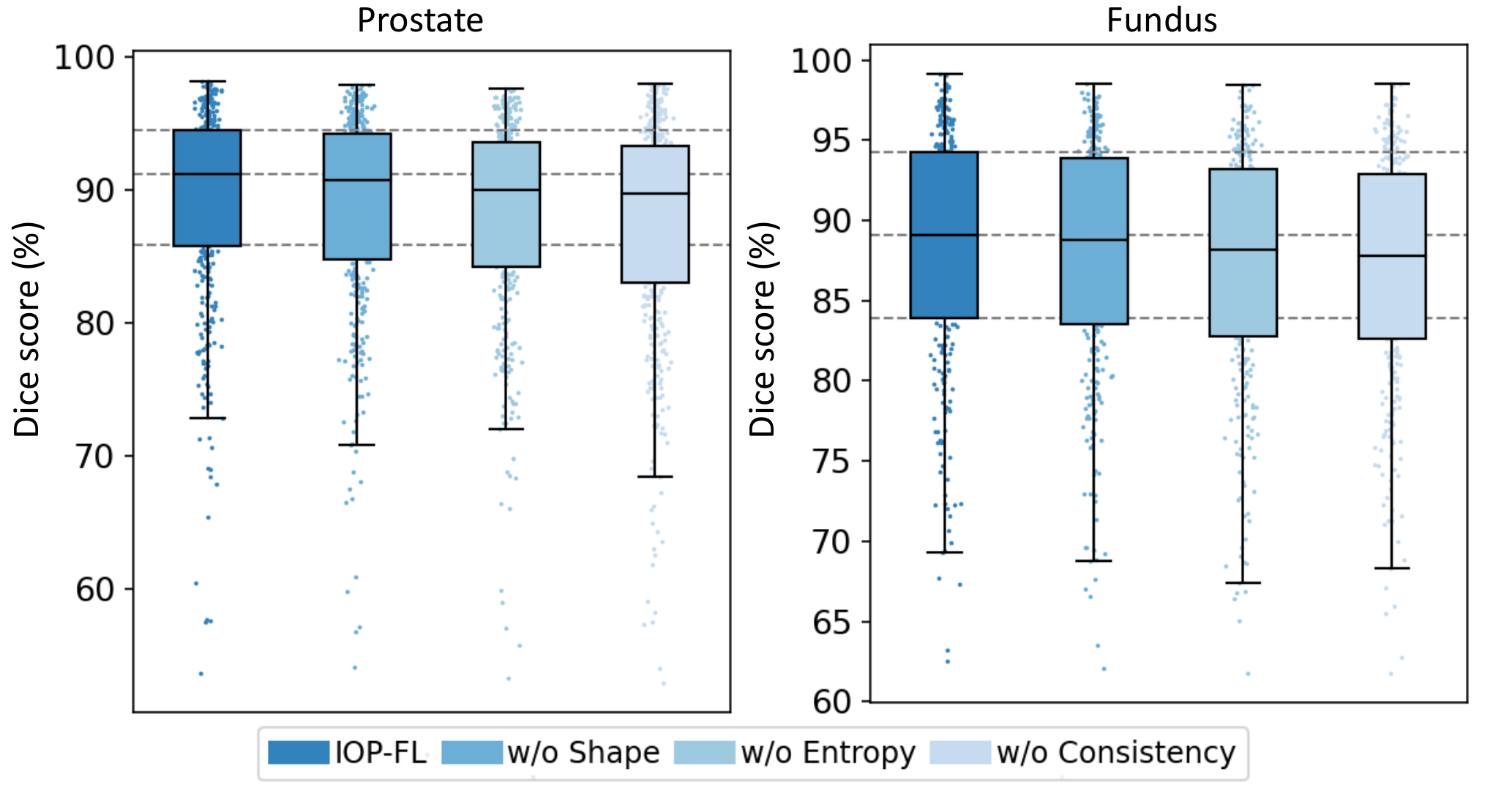}
\caption{{Performance of our method on the outside client when different loss items are used in the test-time routing scheme.}}
\label{fig:loss_ablation}
\end{figure}

\subsection{Ablation Studies}
We further conduct ablation studies to investigate the key properties of our IOP-FL framework. For inside personalization, we analyze: 1) convergence with different local training epochs; and 2) effect of accumulation rate $\tau$. For outside personalization, we analyze: 1) scalability regarding participating number of inside clients; 2) how is the performance of our test-time routing regarding time cost; {3) how is the performance regarding different routing strategies; 4) what is the contribution of each test-time loss item, 5) how the model weight} {variation affects test-time routing, and 6) robustness to model inversion attacks when setting up the outside personalization.}

\subsubsection{Inside Personalization}
\textbf{Convergence with different local epochs:}
Different aggregating frequencies may affect the learning behavior, here we analyze the training performance curve of our IOP-FL compared with different personalized FL methods. In Fig.~\ref{fig:comm_rounds}, we explore different local training epoch $E=1,2,4,8,10$ with 100 communication rounds. It can be observed that the training curve of our IOP-FL increases faster than others, indicating a faster convergence rate of our method. And at the late stage of training, especially for the small local epoch number, the curve of our method goes smoother, demonstrating the benefits of gradient information accumulation to stabilize the local adapted model training.

\textbf{Effects of the accumulation rate:}
We further study how the accumulation rate $\tau$ affects the performance of our method. The local adapted model focus more on current information with a higher rate and a lower one pays more attention to historical information. We draw the Dice scores variation by changing $\tau=\{0.1,0.3,0.5,0.7,0.9\}$, as shown in Fig.~\ref{fig:vary_tau}. For inside clients, $\tau \geq 0.5$ has better results on both datasets with smaller standard deviations. For outside clients, the performance also shows an increasing trend with higher accumulation rate. We further draw the $\tau=1.0$ with dashed lines, the results indicate that totally discarding all previous history may not affect the performance of inside personalization, but it leads to a large performance drop for outside personalization. Typically, our approach can have a good performance with the $\tau$ of $0.9$ for both inside and outside clients.

\subsubsection{Outside Personalization}
\textbf{Outside performance regarding different inside client numbers:}
We analyze how the performance of our method on the outside client would change when different numbers of inside clients participate in FL.
Fig.~\ref{fig:inf_acc_scalable} (a) shows the results on one prostate MRI outside client with the inside client number from 1 to $K\!\!-\!\!1$.
We see that the performance can be improved with more clients being involved, indicating the necessity of using FL to aggregate more data with different distributions. 
Furthermore, compared with FedAvg, the improvement of our method has an increasing trend, i.e., a higher performance gain can be obtained with more clients involved. This demonstrates the efficacy of our test-time routing to exploit a diverse and informative routing space represented by a more number of local models.

\textbf{Inference speed-performance relation study:}
For test-time personalization, it is very important to achieve a quick inference speed as well as maintain a high performance. To this end, we explore the relationship between the test performance on outside clients and the inference time of our method. We use the prostate datasets and plot the overall performance curve by leaving each one of six clients as the outside client, and adding the standard deviation range in Fig.~\ref{fig:inf_acc_scalable} (b). The x-axis denotes the average time to process a single sample. We can see that with more time consumed, i.e., training more iterations with our test-time routing scheme, the performance quickly increases and then saturates. The upper bound of the shaded areas shows that for some outside clients, our approach can fit their distributions at the very beginning with high and stable performance. As for the lower bound, the test Dice score quickly increases and reaches a point near the highest performance with less than one second on each sample, showing the effectiveness of our approach to adapt to various outside clients with less time consumed.
\begin{table}[t]
    \renewcommand\arraystretch{1.2}
    \centering
        \caption{\small{{Comparison of global and layer manner combination on the prostate MRI segmentation.}}}
        \centering
        \vspace{-2mm}
        {
        \scalebox{1.05}{
        \setlength\tabcolsep{3pt}
        \begin{tabular}{c|cccccc |c}
            \toprule \\ [-3.7ex]
             Method&A & B & C & D & E & F & Avg.\\
            \hline
           IOP-FL (Global)
           &88.45&	84.27	&85.02	&86.39	&81.25&	89.77	&85.86
           \\ [-0.5ex]
           IOP-FL (Layer)
           &\textbf{90.52}	&\textbf{90.52}	&\textbf{88.32}&	\textbf{89.39}&	\textbf{84.33}&	\textbf{92.61}&	\textbf{89.28}
           \\ [-0.5ex]
           \bottomrule
        \end{tabular}
    }}
    \label{table:global_layer_prostate}
\end{table}

{
\textbf{Different routing strategies study:}
During our test-time routing, we choose a layer-wise manner to combine training obtained models towards generating the outside personalized model. Using layer-wise combination allows us to adjust the re-weighting coefficient for each layer of each model, which is more flexible than using one global coefficient. Furthermore, it provides a stronger capability to fit the unseen test data than global coefficients. To demonstrate the benefits, we compare the performance on outside clients using global weight combination and layer weight combination on the prostate dataset. As shown in Table~\ref{table:global_layer_prostate},
the layer-wise combination presents significant performance improvements, which our discussion that it is more flexible and shows stronger capability to fit the unseen test data than global combination.}

{\textbf{Contribution of each loss item for test-time routing:}
We validate the effects of key designs for our test-time routing, including the consistency loss, unsupervised shape constraint, and prediction entropy.
We report the results by removing either one component, as shown in Fig.~\ref{fig:loss_ablation}.
From the results can be seen that removing either loss term will decrease the final performance on both segmentation tasks. The trend is clearly shown if we compare the median and quantiles. This consistent observation shows the contributions of our key designs. That is, the consistency loss helps the outside personalized model to fit the overall context distributions, the shape constraint focuses more on the detailed structural information, and the entropy helps minimize the prediction uncertainty.}

\begin{figure}[t]
\centering
\includegraphics[width=0.99\columnwidth]{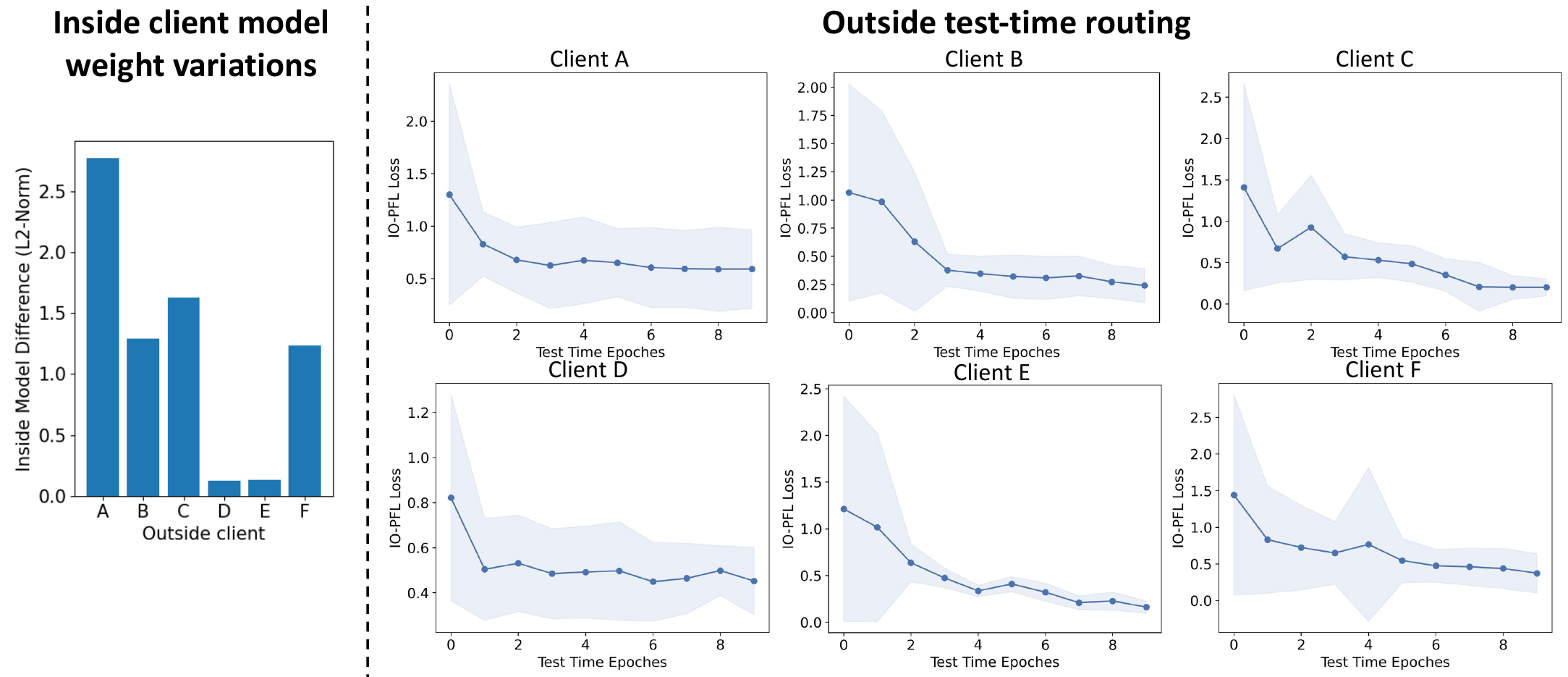}
\vspace{-5mm}
\caption{{Inside client model weight variations and test-time loss curves on the prostate MRI dataset.}}
\label{fig:prostate_variation}
\end{figure}

{\textbf{Model weight variations during test-time routing:}
We} {further investigate the model weight variations during inside training, and how it affects the test-time routing. We first measure the model variation by calculating the average L2-norm of each model layer weight differences between local models and the global model. We perform the study on the prostate dataset and the variation results are shown in the left part in Fig.~\ref{fig:prostate_variation}.
It presents different variation values when choosing a different client as the outside client, indicating that the inside client models vary when doing the outside personalization.
Next, we demonstrate that our test-time optimization method minimizes the test-time loss, which helps the outside personalized model converges under different model variations. We reported the test-time loss value (mean and standard deviation of all test samples) with different epochs. From the figure it can be observed that, in our scenario, the loss value for each outside client can decrease as test-time optimization goes on.}

\begin{figure}[htbp]
\centering
\includegraphics[width=0.99\columnwidth]{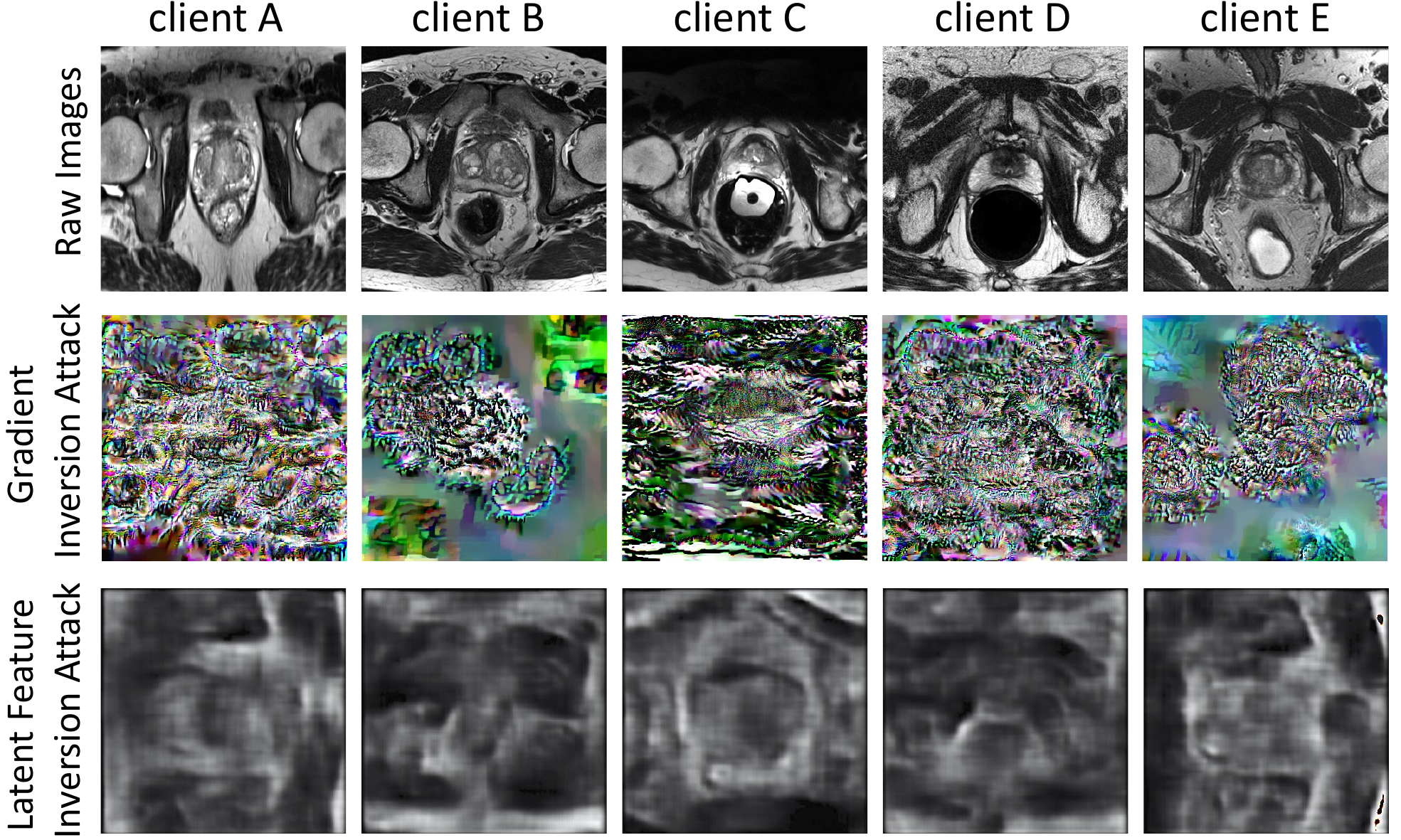}
\caption{Visualization of inversion attacks to reconstruct samples.}
\vspace{-3mm}
\label{fig:attack}
\end{figure}

\black{
\textbf{Robustness to model inversion attacks:}
When performing the outside personalization, the trained models need to be transferred to outside clients, which may raise risks for the model inversion attack. However, with only models available, the model inversion attack from outside clients is difficult to perform if no additional information (e.g., gradient) is given~\cite{fredrikson2015model,zhang2020secret}. Here we investigate two kinds of attacks, the gradient inversion attack~\cite{geiping2020inverting} which utilizes gradients information of training samples to reconstruct original images, and the latent feature inversion attack~\cite{subbanna2021analysis} which requires the latent feature can be accessed by the unseen client. As shown in Fig.~\ref{fig:attack}, even given the gradient or latent feature information, the original data can hardly be perfectly reconstructed.}

\section{Discussions}
With the presence of various distributions of clients in FL, the global model may suffer different degrees of performance degeneration on each local client. This problem becomes even more severe on outside testing clients due to the unseen data distribution shift. To optimize the model prediction accuracy on each individual client, it is important to perform the model personalization, especially for critical medical applications with a low error tolerance~\cite{liu2022medical}. 
{Previous methods have focused on training personalized models for internal clients or adapting/fine-tuning the model for external clients, resulting in separate personalization for inside and outside clients. However, unlike traditional training paradigms, FL can generate multiple models from different clients, providing a new way to create a unified personalization framework.}

{This paper presents a comprehensive strategy to perform model personalization for both inside and outside clients. Specifically, we propose a lightweight gradient-based method to calculate personalized models for inside clients. Our approach derives from the similarity constraints between the pure local model and the global model~\cite{fedprox,moon,pfedme}, but is designed in a more lightweight way by mixing local and global gradients. For outside clients, we propose a novel test-time routing scheme to incorporate all existing models and generate the outside personalized model. Our experiments have shown that simply averaging models predictions or using conventional model ensembles cannot yield satisfactory performance, thus failing to fully exploit the various training-obtained models. Our scheme aims to calculate the outside personalized model by optimizing the re-weight coefficients to aggregate existing models. The coefficients are updated to make the target combined model fit the test data distribution, while the original training knowledge in all internal models is well preserved. With the correct guidance towards fitting new data distribution, the training efforts can further contribute to external clients testing, enhancing the overall applicability of FL.}

{We have demonstrated the applicability of our method in comparison with different methods in Table~\ref{table:inside_all} and Table~\ref{table:outside_all}. Previous methods tackle the inside or outside personalization separately, using very different methods, while our framework solves this problem uniformly for both sides. One concern may lie in the privacy leakage when transferring of all training-obtained models to the unseen client to conduct the personalization. We have investigated two kinds of attacks to show that the original data is hardly to be perfectly reconstructed with the model parameters only. However, the model sharing procedure introduces extra communication cost, and it may require more computing resources to optimize the routing matrix as the number of training clients increases.} Another limitation in our current work is that our shape constraint is designed as prior information to guide the test-time routing on data with elliptical shape, for data with irregular shapes, our shape constraint may provide wrong prior information, which needs to be carefully considered.

Model adaptation becomes necessary when faced with distribution shifts. However, existing methods for adaptation typically require either labeled data or access to source data. For instance, transfer learning by fine-tuning necessitates target labels to retrain the model, while domain adaptation methods require both source and target data to train the model with a cross-domain loss. In real-world scenarios, it may not always be feasible to reprocess source data due to computational constraints, or target labels may not be available for fine-tuning during testing. In some cases, the model may even be deployed without access to source data. Our test-time personalization approach requires only model parameters and unlabeled target data. In comparison to methods that demand source data or target labels, the extra computational cost is reasonable for practical real-world usage. We have also analyzed the inference time cost of our method. By adopting our approach, model performance rapidly improves and stabilizes within one second, demonstrating its efficacy in adapting to diverse external clients while consuming less time. In general, our inside personalization approach and the idea of test-time routing are agnostic to the model architecture. In this work, we mainly consider the segmentation tasks, while this framework can also be applied to other tasks (e.g., diagnosis or detection) via specific changes on the test-time loss design. The exploration of different tasks can further enhance the applicability of our proposed framework, which will be very promising as our future works. In future work, a promising extension of our method is exploring metrics to evaluate the representation capability of internal models, such that the test-time personalization can be achieved by only transferring and combining representative models. This can largely reduce the cost during the model deployment phase for outside clients, especially for the large-scale FL system.

\section{Conclusion}
In this work, we present a novel unified framework, called IOP-FL, to achieve model personalization of each individual client for both inside and outside the federation. In our framework, we propose the local adapted model to serve as the inside personalization and then further utilize the inside personalized models and the global model to perform a newly designed test-time routing toward generating a new outside personalized model. We have conducted comprehensive experiments on two medical image segmentation tasks to demonstrate the effectiveness of our method, and we also performed extensive analytical studies to investigate the key properties of our inside and outside personalization. With the current research trend on applying FL on larger-scale data and becoming closer to real-world deployment, our proposed personalized FL approach has the potential to simultaneously satisfy needs from different clients in practice.
\textcolor{black}{\bibliography{ref.bib}}

\begin{thebibliography}{10}
\providecommand{\url}[1]{#1}
\csname url@samestyle\endcsname
\providecommand{\newblock}{\relax}
\providecommand{\bibinfo}[2]{#2}
\providecommand{\BIBentrySTDinterwordspacing}{\spaceskip=0pt\relax}
\providecommand{\BIBentryALTinterwordstretchfactor}{4}
\providecommand{\BIBentryALTinterwordspacing}{\spaceskip=\fontdimen2\font plus
\BIBentryALTinterwordstretchfactor\fontdimen3\font minus
  \fontdimen4\font\relax}
\providecommand{\BIBforeignlanguage}[2]{{%
\expandafter\ifx\csname l@#1\endcsname\relax
\typeout{** WARNING: IEEEtran.bst: No hyphenation pattern has been}%
\typeout{** loaded for the language `#1'. Using the pattern for}%
\typeout{** the default language instead.}%
\else
\language=\csname l@#1\endcsname
\fi
#2}}
\providecommand{\BIBdecl}{\relax}
\BIBdecl

\bibitem{rieke2020future}
N.~Rieke, J.~Hancox, W.~Li, F.~Milletari, H.~R. Roth, S.~Albarqouni, S.~Bakas,
  M.~N. Galtier, B.~A. Landman \emph{et~al.}, ``The future of digital health
  with federated learning,'' \emph{NPJ digital medicine}, 2020.

\bibitem{silva2019federated}
S.~Silva, B.~A. Gutman, E.~Romero, P.~M. Thompson \emph{et~al.}, ``Federated
  learning in distributed medical databases: Meta-analysis of large-scale
  subcortical brain data,'' in \emph{ISBI}.\hskip 1em plus 0.5em minus
  0.4em\relax IEEE, 2019.

\bibitem{sheller2020federated}
M.~J. Sheller, B.~Edwards, G.~A. Reina, J.~Martin, S.~Pati, A.~Kotrotsou,
  M.~Milchenko, W.~Xu, D.~Marcus \emph{et~al.}, ``Federated learning in
  medicine: facilitating multi-institutional collaborations without sharing
  patient data,'' \emph{Scientific reports}, vol.~10, no.~1, pp. 1--12, 2020.

\bibitem{roth2020federated}
H.~R. Roth, K.~Chang, P.~Singh, N.~Neumark, W.~Li, V.~Gupta, S.~Gupta, L.~Qu,
  A.~Ihsani, B.~C. Bizzo \emph{et~al.}, ``Federated learning for breast density
  classification: A real-world implementation,'' in \emph{DCL, MICCAI
  Workshop}.\hskip 1em plus 0.5em minus 0.4em\relax Springer, 2020.

\bibitem{fedsim}
D.~Li, A.~Kar, N.~Ravikumar, A.~F. Frangi, and S.~Fidler, ``Federated
  simulation for medical imaging,'' in \emph{MICCAI}, 2020.

\bibitem{li2020multi}
X.~Li, Y.~Gu, N.~Dvornek, L.~H. Staib, P.~Ventola, and J.~S. Duncan,
  ``Multi-site fmri analysis using privacy-preserving federated learning and
  domain adaptation: Abide results,'' \emph{MedIA}, vol.~65, p. 101765, 2020.

\bibitem{yeganeh2020inverse}
Y.~Yeganeh, A.~Farshad, N.~Navab, and S.~Albarqouni, ``Inverse distance
  aggregation for federated learning with non-iid data,'' in \emph{DCL, MICCAI
  Workshop}.\hskip 1em plus 0.5em minus 0.4em\relax Springer, 2020.

\bibitem{dayan2021federated}
I.~Dayan, H.~R. Roth, A.~Zhong, A.~Harouni, A.~Gentili, A.~Z. Abidin, A.~Liu,
  A.~B. Costa, B.~J. Wood, C.-S. Tsai \emph{et~al.}, ``Federated learning for
  predicting clinical outcomes in patients with covid-19,'' \emph{Nature
  medicine}, vol.~27, no.~10, pp. 1735--1743, 2021.

\bibitem{ju2020federated}
C.~Ju, D.~Gao, R.~Mane, B.~Tan, Y.~Liu, and C.~Guan, ``Federated transfer
  learning for eeg signal classification,'' in \emph{EMBC}.\hskip 1em plus
  0.5em minus 0.4em\relax IEEE, 2020.

\bibitem{dou2021federated}
Q.~Dou, T.~Y. So, M.~Jiang, Q.~Liu, V.~Vardhanabhuti, G.~Kaissis, Z.~Li, W.~Si,
  H.~H. Lee, K.~Yu \emph{et~al.}, ``Federated deep learning for detecting
  covid-19 lung abnormalities in ct: a privacy-preserving multinational
  validation study,'' \emph{NPJ digital medicine}, vol.~4, no.~1, pp. 1--11,
  2021.

\bibitem{jiang2021harmofl}
M.~Jiang, Z.~Wang, and Q.~Dou, ``Harmofl: Harmonizing local and global drifts
  in federated learning on heterogeneous medical images,'' \emph{AAAI
  Conference on Artificial Intelligence}, 2022.

\bibitem{liu2021federated}
Q.~Liu, H.~Yang, Q.~Dou, and P.-A. Heng, ``Federated semi-supervised medical
  image classification via inter-client relation matching,'' in
  \emph{MICCAI}.\hskip 1em plus 0.5em minus 0.4em\relax Springer, 2021, pp.
  325--335.

\bibitem{tan2022toward}
A.~Z. Tan, H.~Yu, L.~Cui, and Q.~Yang, ``Toward personalized federated
  learning,'' \emph{IEEE TNNLS}, 2022.

\bibitem{wu2020personalized}
Q.~Wu, K.~He, and X.~Chen, ``Personalized federated learning for intelligent
  iot applications: A cloud-edge based framework,'' \emph{IEEE Open Journal of
  the Computer Society}, vol.~1, pp. 35--44, 2020.

\bibitem{kulkarni2020survey}
V.~Kulkarni, M.~Kulkarni, and A.~Pant, ``Survey of personalization techniques
  for federated learning,'' in \emph{WorldS4}.\hskip 1em plus 0.5em minus
  0.4em\relax IEEE, 2020.

\bibitem{kairouz2021advances}
P.~Kairouz, H.~B. McMahan, B.~Avent, A.~Bellet, M.~Bennis, A.~N. Bhagoji,
  K.~Bonawitz, Z.~Charles, G.~Cormode, R.~Cummings \emph{et~al.}, ``Advances
  and open problems in federated learning,'' \emph{Foundations and
  Trends{\textregistered} in Machine Learning}, vol.~14, no. 1--2, pp. 1--210,
  2021.

\bibitem{hsieh2020non}
K.~Hsieh, A.~Phanishayee \emph{et~al.}, ``The non-iid data quagmire of
  decentralized machine learning,'' in \emph{ICML}.\hskip 1em plus 0.5em minus
  0.4em\relax PMLR, 2020.

\bibitem{xu2021federated}
J.~Xu, B.~S. Glicksberg, C.~Su, P.~Walker, J.~Bian, and F.~Wang, ``Federated
  learning for healthcare informatics,'' \emph{Journal of Healthcare
  Informatics Research}, vol.~5, no.~1, pp. 1--19, 2021.

\bibitem{fedbn}
X.~Li, M.~Jiang, X.~Zhang, M.~Kamp, and Q.~Dou, ``Fed{BN}: Federated learning
  on non-{IID} features via local batch normalization,'' in \emph{International
  Conference on Learning Representations}, 2021.

\bibitem{dinh2021fedu}
C.~T. Dinh, T.~T. Vu, N.~H. Tran, M.~N. Dao, and H.~Zhang, ``Fedu: A unified
  framework for federated multi-task learning with laplacian regularization,''
  \emph{arXiv preprint arXiv:2102.07148}, 2021.

\bibitem{smith2017federated}
V.~Smith, C.-K. Chiang, M.~Sanjabi, and A.~Talwalkar, ``Federated multi-task
  learning,'' \emph{NeurIPS}, 2017.

\bibitem{pfedme}
C.~T~Dinh, N.~Tran, and T.~D. Nguyen, ``Personalized federated learning with
  moreau envelopes,'' \emph{NeurIPS}, vol.~33, 2020.

\bibitem{fallah2020personalized}
A.~Fallah, A.~Mokhtari, and A.~Ozdaglar, ``Personalized federated learning: A
  meta-learning approach,'' \emph{NeurIPS}, 2020.

\bibitem{fedbabu}
J.~Oh, S.~Kim, and S.-Y. Yun, ``Fedbabu: Towards enhanced representation for
  federated image classification,'' \emph{ICLR}, 2022.

\bibitem{fedrep}
L.~Collins, H.~Hassani, A.~Mokhtari, and S.~Shakkottai, ``Exploiting shared
  representations for personalized federated learning,'' in \emph{International
  Conference on Machine Learning}.\hskip 1em plus 0.5em minus 0.4em\relax PMLR,
  2021.

\bibitem{fedper}
M.~G. Arivazhagan, V.~Aggarwal, A.~K. Singh, and S.~Choudhary, ``Federated
  learning with personalization layers,'' \emph{arXiv preprint
  arXiv:1912.00818}, 2019.

\bibitem{ttt}
Y.~Sun, X.~Wang, Z.~Liu, J.~Miller, A.~Efros, and M.~Hardt, ``Test-time
  training with self-supervision for generalization under distribution
  shifts,'' in \emph{ICML}.\hskip 1em plus 0.5em minus 0.4em\relax PMLR, 2020,
  pp. 9229--9248.

\bibitem{wang2021tent}
D.~Wang, E.~Shelhamer, S.~Liu \emph{et~al.}, ``Tent: Fully test-time adaptation
  by entropy minimization,'' in \emph{ICLR}, 2021.

\bibitem{wang2019federated}
K.~Wang, R.~Mathews, C.~Kiddon, H.~Eichner, F.~Beaufays, and D.~Ramage,
  ``Federated evaluation of on-device personalization,'' \emph{arXiv preprint
  arXiv:1910.10252}, 2019.

\bibitem{ghosh2020efficient}
A.~Ghosh, J.~Chung, D.~Yin, and K.~Ramchandran, ``An efficient framework for
  clustered federated learning,'' \emph{NeurIPS}, 2020.

\bibitem{sattler2020clustered}
F.~Sattler, K.-R. M{\"u}ller, and W.~Samek, ``Clustered federated learning:
  Model-agnostic distributed multitask optimization under privacy
  constraints,'' \emph{IEEE TNNLS}, 2020.

\bibitem{yu2020salvaging}
T.~Yu, E.~Bagdasaryan, and V.~Shmatikov, ``Salvaging federated learning by
  local adaptation,'' \emph{arXiv preprint arXiv:2002.04758}, 2020.

\bibitem{li2019fedmd}
D.~Li and J.~Wang, ``Fedmd: Heterogenous federated learning via model
  distillation,'' \emph{NeurIPS 2019 Workshop on Federated Learning for Data
  Privacy and Confidentiality}, 2019.

\bibitem{roth2021federated}
H.~R. Roth, D.~Yang, W.~Li, A.~Myronenko, W.~Zhu, Z.~Xu \emph{et~al.},
  ``Federated whole prostate segmentation in mri with personalized neural
  architectures,'' in \emph{MICCAI}.\hskip 1em plus 0.5em minus 0.4em\relax
  Springer, 2021, pp. 357--366.

\bibitem{chen2021personalized}
Z.~Chen, M.~Zhu, C.~Yang, and Y.~Yuan, ``Personalized retrogress-resilient
  framework for real-world medical federated learning,'' in
  \emph{MICCAI}.\hskip 1em plus 0.5em minus 0.4em\relax Springer, 2021, pp.
  347--356.

\bibitem{liu2021feddg}
Q.~Liu, C.~Chen, J.~Qin, Q.~Dou, and P.-A. Heng, ``Feddg: Federated domain
  generalization on medical image segmentation via episodic learning in
  continuous frequency space,'' \emph{CVPR}, 2021.

\bibitem{fedadg}
L.~Zhang, X.~Lei, Y.~Shi, H.~Huang, and C.~Chen, ``Federated learning with
  domain generalization,'' \emph{arXiv preprint arXiv:2111.10487}, 2021.

\bibitem{fedprox}
T.~Li, A.~K. Sahu, M.~Zaheer, M.~Sanjabi \emph{et~al.}, ``Federated
  optimization in heterogeneous networks,'' in \emph{MLSys}, 2020.

\bibitem{hanzely2020federated}
F.~Hanzely and P.~Richt{\'a}rik, ``Federated learning of a mixture of global
  and local models,'' \emph{arXiv preprint arXiv:2002.05516}, 2020.

\bibitem{mcmahan2017communication}
B.~McMahan, E.~Moore, D.~Ramage, S.~Hampson, and B.~A. y~Arcas,
  ``Communication-efficient learning of deep networks from decentralized
  data,'' in \emph{AISTATS}, 2017, pp. 1273--1282.

\bibitem{unet}
O.~Ronneberger, P.~Fischer, and T.~Brox, ``U-net: Convolutional networks for
  biomedical image segmentation,'' in \emph{MICCAI}.\hskip 1em plus 0.5em minus
  0.4em\relax Springer, 2015.

\bibitem{liu2020ms}
Q.~Liu, Q.~Dou, L.~Yu, and P.~A. Heng, ``Ms-net: Multi-site network for
  improving prostate segmentation with heterogeneous mri data,'' \emph{IEEE
  Transactions on Medical Imaging}, 2020.

\bibitem{prostate2014evaluation}
G.~Litjens, R.~Toth, W.~van~de Ven, C.~Hoeks, S.~Kerkstra, B.~van Ginneken,
  G.~Vincent, G.~Guillard, N.~Birbeck, J.~Zhang \emph{et~al.}, ``Evaluation of
  prostate segmentation algorithms for mri: the promise12 challenge,''
  \emph{Medical image analysis}, vol.~18, no.~2, pp. 359--373, 2014.

\bibitem{prostate2015computer}
G.~Lema{\^\i}tre, R.~Mart{\'\i}, J.~Freixenet, J.~C. Vilanova, P.~M. Walker,
  and F.~Meriaudeau, ``Computer-aided detection and diagnosis for prostate
  cancer based on mono and multi-parametric mri: a review,'' \emph{Computers in
  biology and medicine}, vol.~60, pp. 8--31, 2015.

\bibitem{isbi}
B.~Nicholas, M.~Anant, H.~Henkjan, F.~John, K.~Justin \emph{et~al.},
  ``Nci-proc. ieee-isbi conf. 2013 challenge: Automated segmentation of
  prostate structures,'' The Cancer Imaging Archive, 2015.

\bibitem{fumero2011fundus}
F.~Fumero, S.~Alay{\'o}n, J.~L. Sanchez, J.~Sigut, and M.~Gonzalez-Hernandez,
  ``Rim-one: An open retinal image database for optic nerve evaluation,'' in
  \emph{CBMS}.\hskip 1em plus 0.5em minus 0.4em\relax IEEE, 2011, pp. 1--6.

\bibitem{sivaswamy2015fundus}
J.~Sivaswamy, S.~Krishnadas, A.~Chakravarty, G.~Joshi, A.~S. Tabish
  \emph{et~al.}, ``A comprehensive retinal image dataset for the assessment of
  glaucoma from the optic nerve head analysis,'' \emph{JSM Biomedical Imaging
  Data Papers}, vol.~2, no.~1, p. 1004, 2015.

\bibitem{orlando2020fundus}
J.~I. Orlando, H.~Fu, J.~B. Breda, K.~van Keer, D.~R. Bathula, A.~Diaz-Pinto,
  R.~Fang, P.-A. Heng, J.~Kim, J.~Lee \emph{et~al.}, ``Refuge challenge: A
  unified framework for evaluating automated methods for glaucoma assessment
  from fundus photographs,'' \emph{Medical image analysis}, 2020.

\bibitem{fredrikson2015model}
M.~Fredrikson \emph{et~al.}, ``Model inversion attacks that exploit confidence
  information and basic countermeasures,'' \emph{ACM CCS}, 2015.

\bibitem{zhang2020secret}
Y.~Zhang \emph{et~al.}, ``The secret revealer: Generative model-inversion
  attacks against deep neural networks,'' \emph{CVPR}, 2020.

\bibitem{geiping2020inverting}
J.~Geiping \emph{et~al.}, ``Inverting gradients--how easy is it to break
  privacy in federated learning?'' \emph{NeurIPS}, 2020.

\bibitem{subbanna2021analysis}
N.~Subbanna \emph{et~al.}, ``An analysis of the vulnerability of two common
  deep learning-based medical image segmentation techniques to model inversion
  attacks,'' \emph{Sensors}, vol.~21, no.~11, p. 3874, 2021.

\bibitem{liu2022medical}
X.~Liu, B.~Glocker, M.~M. McCradden, M.~Ghassemi, A.~K. Denniston, and
  L.~Oakden-Rayner, ``The medical algorithmic audit,'' \emph{The Lancet Digital
  Health}, 2022.

\bibitem{moon}
Q.~Li, B.~He, and D.~Song, ``Model-contrastive federated learning,'' in
  \emph{CVPR}, 2021, pp. 10\,713--10\,722.

\end{thebibliography}
\bibliographystyle{IEEEtran}
\end{document}